\documentclass[onecolumn]{mn2e}  
\usepackage[dvips]{graphics}
\usepackage{epsfig}
\usepackage{epsf}
\usepackage[figuresleft]{rotating}
\begin{document}
%
\newcommand{\getinptfile}
{\typein[\inptfile]{Input file name}
\input{\inptfile}}

\newcommand{\mysummary}[2]{\noi {\bf SUMMARY}#1 \\ \noi \sl #2 \\ \capline 
	\hspace{-.13in} \raisebox{.0in}{$\sqcap$} \rm }  
\newcommand{\mycaption}[2]{\caption[#1]{\footnotesize #2}} 
\newcommand{\capline}{\mbox{}\hrulefill}
\newcommand{\mysection}[2]{ 
\section{\uppercase{\normalsize{\bf #1}}} \def\junksec{{#2}} } %
\newcommand{\mychapter}[2]{ \chapter{#1} \def\junkchap{{#2}}  
\def\thesection{\arabic{chapter}.\arabic{section}}
\def\thesubsection{\thesection.\arabic{subsection}}
\def\thesubsubsection{\thesubsection.\arabic{subsubsection}}
\def\theequation{\arabic{chapter}.\arabic{equation}}
\def\thefigure{\arabic{chapter}.\arabic{figure}}
\def\thetable{\arabic{chapter}.\arabic{table}}
}
\newcommand{\mysubsection}[2]{ \subsection{#1} \def\junksubsec{{#2}} }
\def\thenote{\addtocounter{footnote}{1}$^{\scriptstyle{\arabic{footnote}}}$ }

\newcommand{\myfm}[1]{\mbox{$#1$}}
\def\spose#1{\hbox to 0pt{#1\hss}}	
\def\ltabout{\mathrel{\spose{\lower 3pt\hbox{$\mathchar"218$}} 
     \raise 2.0pt\hbox{$\mathchar"13C$}}}
\def\gtabout{\mathrel{\spose{\lower 3pt\hbox{$\mathchar"218$}}
     \raise 2.0pt\hbox{$\mathchar"13E$}}}
\newcommand{\ltsim}{\raisebox{-0.5ex}{$\;\stackrel{<}{\scriptstyle \backslash}\;$}}
\newcommand{\simlt}{\ltsim}
\newcommand{\simgt}{\gtsim}
%
\newcommand{\unit}[1]{\ifmmode \:\mbox{\rm #1}\else \mbox{#1}\fi}
\newcommand{\ze}{\ifmmode \mbox{z=0}\else \mbox{$z=0$ }\fi }

%
\newcommand{\boldv}[1]{\ifmmode \mbox{\boldmath $ #1$} \else 
 \mbox{\boldmath $#1$} \fi}
%
\renewcommand{\sb}[1]{_{\rm #1}}%
\newcommand{\expec}[1]{\myfm{\left\langle #1 \right\rangle}}
\newcommand{\mone}{\myfm{^{-1}}}
\newcommand{\half}{\myfm{\frac{1}{2}}}
\newcommand{\nth}[1]{\myfm{#1^{\small th}}}
\newcommand{\ten}[1]{\myfm{\times 10^{#1}}}
\newcommand{\abs}[1]{\mid\!\! #1 \!\!\mid}
\newcommand{\as}{a_{\ast}}
\newcommand{\asr}{(a_{\ast}^{2}-R_{\ast}^{2})}
\newcommand{\bvm}{\bv{m}}
\newcommand{\calf}{{\cal F}}
\newcommand{\calI}{{\cal I}}
\newcommand{\calm}{{v/c}}
\newcommand{\calminf}{{(v/c)_{\infty}}}
\newcommand{\calQ}{{\cal Q}}
\newcommand{\calR}{{\cal R}}
\newcommand{\calw}{{\it W}}
\newcommand{\co}{c_{o}}
\newcommand{\cs}{C_{\sigma}}
\newcommand{\cst}{\tilde{C}_{\sigma}}
\newcommand{\cv}{C_{v}}
\def\dbar{{\mathchar '26\mkern-9mud}}	
\newcommand{\deldelr}{\frac{\partial}{\partial r}}
\newcommand{\deldelR}{\frac{\partial}{\partial R}}
\newcommand{\deldeltheta}{\frac{\partial}{\partial \theta} }
\newcommand{\deldelphi}{\frac{\partial}{\partial \phi} }
\newcommand{\ddotrc}{\ddot{R}_{c}}
\newcommand{\ddotxc}{\ddot{x}_{c}}
\newcommand{\dotrc}{\dot{R}_{c}}
\newcommand{\dotxc}{\dot{x}_{c}}
\newcommand{\Estar}{E_{\ast}}
\newcommand{\grpsi}{\Psi_{\ast}^{\prime}}
\newcommand{\kboltz}{k_{\beta}}
\newcommand{\levi}[1]{\epsilon_{#1}}
\newcommand{\limaso}[1]{$#1 ( a_{\ast}\rightarrow 0)\ $}
\newcommand{\limasinfty}[1]{$#1 ( a_{\ast}\rightarrow \infty)\ $}
\newcommand{\limrinfty}[1]{$#1 ( R\rightarrow \infty,t)\ $}
\newcommand{\limro}[1]{$#1 ( R\rightarrow 0,t)\ $}
\newcommand{\limrso}[1]{$#1 (R_{\ast}\rightarrow 0)\ $}
\newcommand{\limxo}[1]{$#1 ( x\rightarrow 0,t)\ $}
\newcommand{\limxso}[1]{$#1 (\xs\rightarrow 0)\ $}
\newcommand{\ls}{l_{\ast}}
\newcommand{\Ls}{L_{\ast}}
\newcommand{\mean}[1]{<#1>}
\newcommand{\ms}{m_{\ast}}
\newcommand{\Ms}{M_{\ast}}
\def\nb{{\sl N}-body }
\def\nbt{{\sf NBODY2} }
\def\nb1{{\sf NBODY1} }
\newcommand{\nuoned}{\nu\sb{1d}}
\newcommand{\ra}{\rightarrow}
\newcommand{\Ra}{\Rightarrow}
\newcommand{\rc}{r_{c} } 
\newcommand{\Rc}{R_{c} } 
\newcommand{\res}[1]{{\rm O}(#1)}
\newcommand{\rnsa}{(r^{2}-a^{2})}
\newcommand{\Rnsa}{(R^{2}-a^{2})}
\newcommand{\rs}{r_{\ast}}
\newcommand{\Rs}{R_{\ast}}
\newcommand{\Rsa}{(R_{\ast}^{2}-a_{\ast}^{2})}
\newcommand{\sa}{\sigma } 
\newcommand{\sac}{\sigma_{c} } 
\newcommand{\sas}{\sigma_{\ast} } 
\newcommand{\sasp}{\sigma^{\prime}_{\ast}}
\newcommand{\saxs}{\sigma_{\ast} } 
\newcommand{\sech}{{\rm sech}}
\newcommand{\tff}{t\sb{ff}} 
\newcommand{\ti}{\tilde}
\newcommand{\trel}{t\sb{rel}}
\newcommand{\ts}{\tilde{\sigma} } 
\newcommand{\tss}{\tilde{\sigma}_{\ast} } 
\newcommand{\vcol}{v\sb{col}}
\newcommand{\vs}{v_{\ast}  } 
\newcommand{\vsp}{v^{\prime}_{\ast}}
\newcommand{\vxs}{v_{\ast}  } 
\newcommand{\xs}{x_{\ast}}
\newcommand{\xc}{x_{c} } 
\newcommand{\xistar}{\xi_{\ast}}
\newcommand{\rmd}{\ifmmode \:\mbox{{\rm d}}\else \mbox{ d}\fi }
\newcommand{\rmD}{\ifmmode \:\mbox{{\rm D}}\else \mbox{ D}\fi }
\newcommand{\valfven}{v_{{\rm Alfv\acute{e}n}}}

%
\newcommand{\noi}{\noindent}
\newcommand{\bc}{boundary condition }
\newcommand{\bcs}{boundary conditions }
\newcommand{\Bcs}{Boundary conditions }
\newcommand{\lhs}{left-hand side }
\newcommand{\rhs}{right-hand side }
\newcommand{\wrt}{with respect to }
\newcommand{\iras}{{\sl IRAS }}
\newcommand{\cobe}{{\sl COBE }}
\newcommand{\Oh}{\myfm{\Omega h}}
%
\newcommand{\etal}{{\em et al.\/ }}
\newcommand{\eg}{{\em e.g.\/ }}
\newcommand{\etc}{{\em etc.\/ }}
\newcommand{\ie}{{\em i.e.\/ }}
\newcommand{\viz}{{\em viz.\/ }}
\newcommand{\cf}{{\em cf.\/ }}
\newcommand{\via}{{\em via\/ }}
\newcommand{\apriori}{{\em a priori\/ }}
\newcommand{\adhoc}{{\em ad hoc\/ }}
\newcommand{\viceversa}{{\em vice versa\/ }}
\newcommand{\versus}{{\em versus\/ }}
\newcommand{\qed}{{\em q.e.d. \/}}
\newcommand{\<}{\thinspace}
%
\newcommand{\km}{\unit{km}}
\newcommand{\kms}{\unit{km~s\mone}}
\newcommand{\kmsa}{\unit{km~s\mone~arcmin}}
\newcommand{\kpc}{\unit{kpc}}
\newcommand{\mpc}{\unit{Mpc}}
\newcommand{\hkpc}{\myfm{h\mone}\kpc}
\newcommand{\hmpc}{\myfm{h\mone}\mpc}
\newcommand{\parsec}{\unit{pc}}
\newcommand{\cm}{\unit{cm}}
\newcommand{\yr}{\unit{yr}}
\newcommand{\au}{\unit{A.U.}}
\newcommand{\AU}{\au}
\newcommand{\gm}{\unit{g}}
\newcommand{\solarm}{\unit{M\sun}}
\newcommand{\Lsun}{\unit{L\sun}}
\newcommand{\Rsun}{\unit{R\sun}}
\newcommand{\seconds}{\unit{s}}
\newcommand{\micro}{\myfm{\mu}}
\newcommand{\Mdot}{\myfm{\dot M}}
%
%
%
\newcommand{\dgr}{\myfm{^\circ} }
\newcommand{\ddgr}{\mbox{\dgr\hskip-0.3em .}}
\newcommand{\mnt}{\mbox{\myfm{'}\hskip-0.3em .}}
\newcommand{\scnd}{\mbox{\myfm{''}\hskip-0.3em .}}
\newcommand{\hr}{\myfm{^{\rm h}}}
\newcommand{\dhr}{\mbox{\hr\hskip-0.3em .}}
%
%
%
%
%
%
%
\newcommand{\refindent}{\par\noindent\hangindent=0.5in\hangafter=1}
\newcommand{\figpar}{\par\noindent\hangindent=0.7in\hangafter=1}
%
%

\newcommand{\mybiblio}{\vspace{1cm}
		       \setcounter{subsection}{0}
		       \addtocounter{section}{1}
		       \def\junksec{References} 
 }

%
%
%

%
%
%
%
%

\newcommand{\vol}[2]{ {\bf#1}, #2}
\newcommand{\jour}[4]{#1. {\it #2\/}, {\bf#3}, #4}
\newcommand{\physrevd}[3]{\jour{#1}{Phys Rev D}{#2}{#3}}
\newcommand{\physrevlett}[3]{\jour{#1}{Phys Rev Lett}{#2}{#3}}
\newcommand{\aaa}[3]{\jour{#1}{A\&A}{#2}{#3}}
\newcommand{\aaarev}[3]{\jour{#1}{A\&A Review}{#2}{#3}}
\newcommand{\aaas}[3]{\jour{#1}{A\&A Supp.}{#2}{#3}}
\newcommand{\aj}[3]{\jour{#1}{AJ}{#2}{#3}}
\newcommand{\apj}[3]{\jour{#1}{ApJ}{#2}{#3}}
\newcommand{\apjl}[3]{\jour{#1}{ApJ Lett.}{#2}{#3}}
\newcommand{\apjs}[3]{\jour{#1}{ApJ Suppl.}{#2}{#3}}
\newcommand{\araa}[3]{\jour{#1}{ARAA}{#2}{#3}}
\newcommand{\mn}[3]{\jour{#1}{MNRAS}{#2}{#3}}
\newcommand{\mnras}{\mn}
\newcommand{\jgeo}[3]{\jour{#1}{Journal of Geophysical Research}{#2}{#3}}
\newcommand{\qjras}[3]{\jour{#1}{QJRAS}{#2}{#3}}
\newcommand{\nat}[3]{\jour{#1}{Nature}{#2}{#3}}
\newcommand{\pasa}[3]{\jour{#1}{PAS Australia}{#2}{#3}}
\newcommand{\pasj}[3]{\jour{#1}{PAS Japan}{#2}{#3}}
\newcommand{\pasp}[3]{\jour{#1}{PAS Pacific}{#2}{#3}}
\newcommand{\rmp}[3]{\jour{#1}{Rev. Mod. Phys.}{#2}{#3}}
\newcommand{\science}[3]{\jour{#1}{Science}{#2}{#3}}
\newcommand{\vistas}[3]{\jour{#1}{Vistas in Astronomy}{#2}{#3}}

%
%
%
\newcommand{\leftb}{<\!\!} \newcommand{\rightb}{\!\!>}
\newcommand{\oversim}[2]{\protect{\mbox{\lower0.5ex\vbox{%
  \baselineskip=0pt\lineskip=0.2ex
  \ialign{$\mathsurround=0pt #1\hfil##\hfil$\crcr#2\crcr\sim\crcr}}}}} 
\newcommand{\simgreat}{\mbox{$\,\mathrel{\mathpalette\oversim>}\,$}} 
\newcommand{\simless} {\mbox{$\,\mathrel{\mathpalette\oversim<}\,$}} 
%
%
%
%
\title[The universal outcome of star formation.]
{On the universal outcome of star-formation: 
Is there a link between stars and brown-dwarfs?}  
   \author[Kroupa, Bouvier, Duch\^ene \& Moraux]
{Pavel Kroupa$^{1,2,4}$, Jerome Bouvier$^1$, Gaspard Duch\^ene$^3$,
   Estelle Moraux$^1$
\\
$^1$Laboratoire d'Astrophysique de l'Observatoire de Grenoble, BP 53,
   F-38041 Grenoble Cedex 9, France \\
$^2$Institut f\"ur Theoretische Physik und Astrophysik der
Universit\"at Kiel, D-24098 Kiel, Germany\\
$^3$Department of Physics and Astronomy, 405 Hilgard Avenue, UCLA, Los
Angeles, CA~90095-1562, USA\\
$^4$ {\it Heisenberg Fellow}
}

\maketitle

\begin{abstract} 
Given the current consensus that stars form from pre-stellar cloud
cores that fragment into small$-N$ groups which decay within a few
$10^4$~yr, and taking the observed properties of about 1~Myr old stars
in the Taurus-Auriga (TA) star-forming region as empirical
constraints, we suggest a model that describes the multiplicity
properties of the disintegrated groups.  This model concisely
describes the outcome of star formation in terms of dynamically
unevolved binary properties.  Two variants of the model are tested
against data on very young stars in Taurus-Auriga (TA) and the Orion
Nebula cluster (ONC) as well as the older Pleiades and the
Galactic-field populations.  The {\it standard model} (SM) assumes
that cloud-core fragmentation only produces stellar systems, while the
{\it standard model with brown dwarfs} (SMwBDs) assumes that
cloud-core fragmentation proceeds down to sub-stellar mass cores.
Brown dwarfs (BDs) enter the SM by being a separate, dynamically
unimportant population.  The models produce a very high initial binary
proportion among stars (SM), and stars and BDs (SMwBDs), and both
reproduce the measured initial mass function (IMF) in TA, the ONC and
the Pleiades as well as the Galactic field.  Concentrating on the
SMwBDs, it is shown that the Briceno et al. result that TA appears to
have produced significantly fewer BDs per star than the ONC is
reproduced almost exactly without calling for a different IMF. The
reason is that star--BD and BD--BD binaries are disrupted in the dense
ONC. The model, however, fails to reproduce the observed star--star
binary period distribution in TA, because it contains too many
star--BD pairs. Also, the SMwBDs leads to too many wide star--BD and
BD--BD systems. This is a problem if most stars form in clusters
because Galactic-field very-low-mass-star and BD binaries have a low
binary fraction and do not contain wide systems. The SM, on the other
hand, finds excellent agreement with the observed mass-ratio and
period distribution among TA and Galactic-field stellar binaries, as
well as the observed stellar period distribution in the ONC and the
Pleiades. The conclusion of this work is therefore that the SM
describes the initial, dynamically unevolved stellar population very
well indeed for a large range of star-forming conditions, suggesting
(1) a remarkable invariance of the star-formation products, and (2)
that BDs (and some very-low-mass stars) need to be added as a separate
population with its own kinematical and binary properties.  This
separate population may vary with star-forming conditions.

{\keywords stars: formation --
stars: low-mass, brown dwarfs -- binaries: general
-- open clusters and associations: general -- Galaxy: stellar content
-- stellar dynamics}

\end{abstract}

\section{Introduction}
\label{sec:intro}

The properties of multiple systems and the shape of the IMF are the
outcome of star formation. By studying these in different environments
it may be possible to unearth variations that are important for
constraining star-formation theory. To achieve this it is necessary to
apply exactly the same methodology to a variety of populations. The
observational side of such a major endeavour is presented by the
consistent work notably by Luhman and collaborators (e.g. Luhman et
al. 2003a) on many very young populations. But it is also necessary to
take into account changes in young stellar populations owing to
stellar-dynamical processes, in order to verify if observed
differences may not merely be due to the dynamical evolution, or even
to uncover true differences in the absence of observed differences.
For example, as has been stressed many times since a first thorough
investigation of such issues by Kroupa, Tout \& Gilmore (1991), {\it a
steep IMF can appear flatter if binary stars are abundant, thus
appearing similar to a truly flat IMF if binary stars are sparse}. It
is worth keeping this bias in mind when investigating possible
variations of primordial stellar populations. This we do here by
concentrating on very recent observational evidence which allow us to
probe star-formation into the BD mass range. 

Briceno et al. (2002) report nine new objects with masses in the range
$0.015-0.1\,M_\odot$ in the TA star forming region. The discovery is
based on a deep optical survey combined with data from the Two-Micron
All-Sky Survey and a follow-up spectroscopic survey of eight groups
known to contain pre-main sequence stellar members. Each of the survey
areas comprises approximately 2.3~pc~$\times$~2.3~pc (for a distance
of 140~pc), and is complete down to masses $m=0.02\,M_\odot$.  They
construct an IMF from the new sample and find a significant deficit of
BDs in TA when compared to the ONC.  {\it This may be the first direct
evidence for a non-universality of the IMF at its low-mass end}.

The assertion by Briceno et al. is very important for star-formation
theory, as it implies that the production of BDs through fragmentation
of the cloud core may be dependent on the physical conditions of the
molecular cloud, given that the cloud cores in which the ONC and the
TA-aggregates formed were very different in density and mass.  Thus,
in TA groups of a few dozen binaries are forming in a volume of about
1~pc$^3$ (Gomez et al. 1993; Hartmann 2002) within which up to~$10^4$
ONC stars may have formed (Kroupa, Aarseth \& Hurley 2001, hereinafter
KAH). As pointed out by Briceno et al., the sense of the discrepancy
between TA and the ONC is expected if the distribution of stellar
masses reflects that of Jeans-unstable fragments. TA should be
producing fewer BDs by direct fragmentation than the ONC, because the
Jeans mass, ${\cal M} \propto \rho^{-1/2}\,T^{3/2}$, so that ${\cal
M}_{\rm TA} > {\cal M}_{\rm ONC}$ since the gas-density is about two
orders of magnitude lower in TA than it was in the ONC, $\rho_{\rm
TA}\approx 10^{-2}\,\rho_{\rm ONC}$, while the temperatures, $T$, are
not likely to have differed by more than a factor of a few.  That BDs
may have a star-like formation history is also tentatively supported
by the mass function (MF) of pre-stellar cores in $\rho$~Oph which
Motte, Andr\'e \& N\'eri (1998) and Bontemps et al. (2001) find to be
indistinguishable from the very-young (class~II) stars and the
Galactic-field IMF in the mass range $0.06-2\,M_\odot$. The
conditions in $\rho$~Oph appear to be such that the observed masses
are consistent with Jeans fragmentation of the molecular cloud down
into the sub-stellar mass range without significant further evolution
of the masses by interactions between proto-stars or competitive
accretion. 

Within the Briceno et al. survey regions the authors find the spatial
distribution of BDs and stars to be indistinguishable and both to show
similar K-band excesses indicating similar accretion-disk processes.
Briceno et al. find {\it no evidence for a different formation
mechanism between BDs and stars}.  This is also supported by the
high-resolution optical spectra of seven low-mass stellar and BD
members of TA obtained by White \& Basri (2003), who find that the
kinematics of the BDs cannot be distinguished from the stellar
motions. They conclude that the BDs form like stars albeit with
smaller disk masses and smaller accretion rates. That is, stars and
BDs should be formed with the same kinematical, spatial and binary
properties.  Given that they find no kinematical differences between
their BDs and the stars {\it they exclude the embryo ejection
hypothesis}.  According to this hypothesis (Reipurth 2000; Boss 2001;
Reipurth \& Clarke 2001; Boss 2002; Bate, Bonnell \& Bromm 2002), BDs
may be unfinished stellar embryos that are expelled from forming
multiple-star systems. The dynamical expulsion leads to relatively
fast moving ($v\simgreat 1$~km/s) BDs with truncated disks (Sterzik \&
Durisen 1998, 2003, hereinafter SD98, SD03; Delgado-Donate, Clarke \&
Bate 2003, hereinafter DCB). The spatial distribution of the so
produced BDs should therefore differ from the distribution of the
stars. Briceno et al. do not observe a difference and they thus also
discard the embryo-ejection hypothesis.

There are thus two possibilities for the origin of BDs that emerge
from the preceding discussion. These can be framed as hypotheses
(\S~\ref{sec:hyp}) that may then be tested against observational data.
To test which hypothesis is consistent with the Briceno et al. data as
well as data that are available for other populations, it is necessary
to construct initial populations and allow these to evolve dynamically
to the required ages.  The theoretical stellar populations are
constructed using a set of minimal assumptions by matching the IMF,
binary-star period, mass-ratio and eccentricity distributions of
T~Tauri stars and late-type main sequence Galactic-field
populations. The simultaneous fit to the pre-main sequence and main
sequence data implies most Galactic-field stars to be born in modest
clusters (Kroupa 1995), which is also arrived at from direct
observational surveys (Lada \& Lada 2003; Carpenter 2000), and an
analysis of the distribution and lifetimes of local Galactic clusters
(Adams \& Myers 2001).  The ``dominant-mode cluster'' contains
initially 200~binaries and has a half-mass radius
$R_{0.5}\approx0.8$~pc.  The so-obtained {\it standard model} (SM) of
star formation (\S~\ref{sec:stand} below) leads to a very good
description of {\it stellar} populations in star clusters and the
Galactic field.

The purpose of this contribution is to test the SM by extending it to
include BDs.  We apply the {\it standard model with BDs} (SMwBDs) to
the TA star-forming region and to the ONC and the Pleiades, and
systematically study the implication the SMwBDs has on the observed
number of BDs and on the orbital-parameter distributions in a variety
of environments such as TA, ONC and Pleiades.  This may shed light on
the origin of BDs. {\it The underlying goal is to seek the simplest
physically-motivated description of the initial population that is
consistent with all available data for different environments. This
will yield useful constraints on the star formation process and will
also provide a realistic input population for extensive star-cluster
modelling.}

The pre-main sequence populations in TA and in the ONC are especially
suited for such a comparison because they offer examples of very
different environments, TA giving birth to groups of a few dozens of
late-type binaries, while the ONC is a post-gas-expulsion cluster
containing a few thousand systems ($=$ single stars plus binaries).
The TA population is composed predominantly of binaries (and some
higher-order multiples) while the ONC has a binary proportion similar
to that in the Galactic field. Stars more massive than about
$1.5\,M_\odot$ do not form in TA because of the limited molecular
cloud masses, while the much more massive ONC precursor gave birth to
a few O~stars.  Both have a similar age (about 1~Myr).

A discussion of state-of-the art theoretical work on star formation
and the line-of-arguments leading to the standard star-formation model
is given in \S~\ref{sec:sf}.  The stellar-dynamical models of the TA
aggregates, the ONC and Pleiades are introduced in \S~\ref{sec:mods}.
Section~\ref{sec:noBDs} models the Briceno et al. data.
\S~\ref{sec:binprop} presents the implications of the SMwBDs on the
period and semi-major axis distributions of stellar and BD binaries,
for standard and non-standard IMFs. The conclusions are given in
\S~\ref{sec:concs}.

Throughout this paper, we refer to the ``stellar IMF'' or simply the
``IMF'' as being the initial mass function of stars, constructed by
counting all stars individually, while the ``initial MF of systems''
or the ``initial system MF'' refers to the distribution of system
masses in the initial population. Likewise, ``stellar MF'' and
``system MF'' refer to possibly evolved mass distributions, and an
``observed MF'' is taken to mean the empirical MF of unresolved
systems. A ``system'' can be a single star or be composed of physically
bound multiple stars.

\section{The outcome of star formation}
\label{sec:sf}

This section addresses the current concensus on star-formation. A
comparison of state-of-the art theoretical work with available
empirical constraints allows us to distill the properties of the
systems that actually form through fragmentation of cloud cores. Thus
we can also infer a useful algorithmic description of the outcome of
this process. This is important because the full problem of cloud
collapse leading to stars cannot be computed fully self-consistently,
but a description of initial populations is nevertheless needed for a
wide variety of astrophysical problems.  The description of this
outcome, referred to as the ``standard model'', details the {\it
characteristics of a dynamically unevolved stellar population}. Such
characteristics are the IMF, the initial binary proportion and the
initial distribution function of angular momenta (periods,
eccentricities and mass-ratios), and can be referred to as being the
{\it dynamical properties of a population}.  The question being
addressed here and in past and future contributions is how the
characteristics of this initial population vary with star forming
conditions. Once these characteristics and their possible variation
with the physical conditions in molecular clouds are known, these can
be used to set-up initial populations for $N$-body computations of
star cluster formation and evolution, as well as to construct entire
Galactic-field populations using the method of ``dynamical population
synthesis'' (the construction of Galactic-field populations from
dispersing star clusters, Kroupa 1995). Knowing how and if at all the
dynamical properties vary would also place important constraints on
the cloud collapse work.
  
\subsection{Cloud-core fragmentation}
\label{sec:clcoll}

In the following we differentiate between a ``cloud core'' that may
form a cluster and a pre-stellar cloud-core or ``kernel'' which only
forms a stellar system (cores of stellar mass with extent of about
0.03~pc, Myers 1998).  It has been realised for some time now that a
collapsing slowly rotating kernel fragments into multiple accreting
hydrostatic cores that initially form a bound system (e.g. Burkert \&
Bodenheimer 1996; Boss 2002). The latest high-resolution
hydrodynamical collapse computations (Bate, Bonnell \& Bromm 2003)
deal with the collapse of a section of a molecular cloud and show that
it develops a distribution of kernels each of which rapidly fragment
into small-$N$ systems that disrupt and merge with the rest of the
system. These results are very encouraging and the latest such
experiment by Bonnell, Bate \& Vine (2003) find a stellar IMF which is
virtually identical to the Galactic-field IMF that has been found to
be rather surprisingly invariant (Kroupa 2002).

A problem faced by all these computations (Kroupa \& Bouvier 2003a,
hereinafter KB1) is that stellar feedback cannot be included so that
it is unknown at this stage how much gas is accreted and how much is
removed again through outflows and photoionisation in the event of OB
stars forming in the vicinity. The fragmentation of magnetised kernels
are reported by Boss (2002) with the result that binaries or multiples
form readily, but the magnetic effects need to be treated in
approximate ways.  In addition, at present the collapse of each kernel
proceeds unhindered and achieves very high densities that lead to
rather violent dynamical evolution of the emerging stellar
system. Such groups decay within about $N$-crossing times (e.g. DCB)
by expelling members stochastically, unless their initial
configuration is hierarchical.  In the presence of feedback the
additional energy input is likely to limit the collapse due to the
higher thermal energy of the gas thus probably leading to more
extended, less violent small-$N$ stellar groups.

Based on the results of available cloud collapse calculations, SD03
elaborate a very useful model which assumes that cloud cores fragment
into $1\le N \le 10$ stars and BDs within a region with half-mass
radius $R_{0.5}\approx125$~AU (from SD98).  The nominal crossing time
is $t_{\rm cross}=(2\,R_{0.5})^{3/2}/(G\,M_{\rm st})^{1/2}$, where
$M_{\rm st}$ is the mass of the group and $G$ is the gravitational
constant. The crossing time is typically $300$~yr. SD03 study the
multiple-star properties of the disintegrated groups by evaluating
their data after $300\,t_{\rm cross}\approx 10^5$~yr. They find good
agreement of their model BDs and stellar systems with those observed
in the Galactic field (their fig.~2, compare with Fig.~\ref{fig:fpm}
below).  

The model appears to be challenged though by the high binary
proportion observed in the TA stellar population. The decay time of
the multiple systems is too short, being typically very much shorter
than $100\,t_{\rm cross} \ll \tau$, to explain the observed high
multiplicity fraction in TA, which is composed mostly of
$\tau\approx1$~Myr old binary systems. This point is shown graphically
by fig.~1 in Reipurth \& Clarke (2001).  Since about 50~per cent of
the ejected single stars have three-dimensional velocities less than
1~pc/Myr (fig.~5 in SD03), {\it a large fraction of single stars would
therefore remain within the observed areas for 1~Myr or longer}, thus
significantly reducing the binary fraction there, in contradiction to
the observations. Also, the stars ejected with $v\simgreat 1$~km/s
would form a halo population of single stars around the stellar
aggregates that is not observed.  The BDs expelled from the SD03
groups have a median velocity of 2pc/Myr implying that the BDs and
stars ought to have well separated in TA, contrary to the conclusions
made by Briceno et al. and others, as discussed in~\S~\ref{sec:intro}.

The results of SD03 are strictly valid only for stellar groups that do
not contain gas, and are as such very important benchmark models in a
regime where the physics is well understood.  The early dynamical
evolution of such groups is likely to be dominated by the gas in the
kernel however, and DCB develop such a small-$N$ model. DCB place
$N=5$ seeds into a non-rotating gaseous core initially in hydrostatic
equilibrium and which comprises 90~per cent of the mass of the
embedded system. The subsequent evolution is governed by competitive
accretion and mutual ejections until either a binary or a long-lived
hierarchical system remains. DCB continue the computations until all
the gas is accreted on the seeds, which is, as they state, not
realistic but which defines another extreme set of models that,
together with the SD03 models, may bracket reality.

The overall result is similar to what SD03 find, namely that the
groups decay rapidly leaving a multiple system.  There are notable
differences though as a result of including gas dynamics.  The
ejection velocities are typically larger, despite the additional
retarding potential given by the mass in gas, as a result of the group
of accreting seeds contracting because of the accretion of low-angular
momentum gas. The shrinking of the groups leads to more energetic
dynamics. Seeds are expelled rapidly, within a few $t_{\rm cross}$, so
that the models predict that mostly BDs are expelled with a median
speed of about 2~pc/Myr (their fig.~8). The remaining seeds that form
hierarchical multiple systems or binaries accrete the rest of the gas
and thus acquire stellar masses. 

The prediction of DCB is thus that BDs have a negligible binary
fraction while the stars have a high multiplicity fraction, close to
70~per cent, with the exact numbers depending on the details of the
convolution between the core and fragment MF (Delgado-Donate, private
communication).  This appears to be consistent with the observations
(Fig.~\ref{fig:fpm} below), but DCB do not state how long-lived their
multiple stars are. Most of the BDs have speeds larger than 1~pc/Myr
and the model produces approximately equal numbers of BDs and stars,
no binaries with mass-ratios smaller than~0.2 and a stellar semi-major
axis distribution which is much narrower than the observed
distribution.  This model thus implies BDs to have different
kinematical and accretion properties than stars, which has been
observed by Briceno et al. (2002) and White \& Basri (2003) not to be
the case.

Given these arguments, it appears necessary to amend the fragmentation
scenario such that the usual outcome is a long-lived hierarchical
stellar system.

\subsection{Two hypotheses}
\label{sec:hyp}

Taking into account the theoretical results and the empirical
constraints discussed above we set-up a model which describes the
outcome of the fragmentation of kernels.  The model is guided more by
the empirical evidence from TA on the size, density, number of stars,
binary-star properties, than by the theoretical results discussed
above since the physics and thus the detailed outcome of fragmentation
of a kernel is not well understood yet.  Thus, while we assume that
pre-stellar cloud-cores fragment into multiple systems, the properties
of the systems must also be consistent with the observational data. A
kernel can thus fragment into a binary or long-lived hierarchical
multiple system. The multiple system can be either a hierarchical
triple or a hierarchical quadruple. The latter can be approximated by
two weakly bound binaries. A physical reason why fragmentation would
not produce chaotic, dynamically violent small-$N$ groups with short
life-times could be the energy input through stellar feedback which
may oppose collapse to high densities and thus limit $N$ per kernel.

From the discussion in~\S~\ref{sec:intro} it follows that there are
two principle possibilities for the origin of BDs.  These can be
framed as hypotheses that may then be tested against observational
data:

\begin{description}
\itemsep=-4mm

\item
{\it Hypothesis~A}: Only stellar systems form from pre-stellar cores,
and BDs and some very-low-mass stars are unfinished embryos expelled
from the kernels. Motivation of this comes from the work of SD03 and
DCB (e.g. their fig.~14) and is consistent with the embryo ejection
hypothesis. Hypothesis~A leads to the ``standard model'' (SM),
according to which BDs are a separate population of mostly single
objects. This additional population is dynamically insignificant
because it contributes less than 5~per cent in mass for usual IMFs
(Kroupa 2002). The formation history of BDs and stars differ
fundamentally. \\

\item
{\it Hypothesis~B}: The fragmentation can also occur in pre-stellar
cores with sub-stellar masses leading to the formation of many BD
binary systems (DCB). This hypothesis essentially states that BDs and
stars form exactly in the same manner, and is consistent with the
above mentioned rejection of the embryo-ejection hypothesis by Briceno
et al. and White \& Basri. Hypothesis~B leads to the ``standard model
with BDs'' (SMwBDs).  It implies that {\it BDs and stars are born with
the same binary and kinematical properties}. The hypothesis assumes
sub-stellar-mass kernels fragment, the parent kernel distribution
thus being assumed to extend well into the sub-stellar regime, as is,
in fact, indicated to be the case in $\rho$~Oph (Motte et al. 1998;
Bontemps et al. 2001). As a result, BDs have a high binary fraction,
as is also emphasised by DCB in their section~7, and there are many
star--BD systems.

\end{description}

Dynamical evolution is known to strongly affect stellar populations
(Kroupa 2001) and observed differences such as in the number of BDs
per star may not trace some intrinsic difference in the initial
populations.  We therefore test both hypotheses taking into account
the dynamical evolution, with an emphasis on hypothesis~B because the
SM is already known to yield an excellent description of stellar
populations in a large variety of systems.  We seek to find possible
discrepancies with empirical data that will allow us to reject or
modify the hypothesis.

\subsection{The standard model of star formation}
\label{sec:stand}

We try here to produce a realistic model of a stellar population and
make it evolve dynamically before comparing it with the observed
properties of TA, ONC and Pleiades. The original version of the
standard model did not include BDs (Kroupa 1995).  This is referred to
as the SM.  A further description of this model is available in KB1,
and Kroupa (1998) discusses the implied properties of runaway stars.
Some of its success is re-iterated here by showing previously
unpublished results, followed by its extension through the inclusion
of BDs, which we refer as the SMwBDs.  Whenever we spell-out
``standard model'' we refer to the general properties of both
variations of this model (SM and SMwBDs).

\subsubsection{The SM (hypothesis~A)}
\label{sec:SM}

A ``standard model'' describing the outcome of low-mass star formation
in terms of an invariant field-like IMF, random pairing of mass from
the IMF to form binaries, a birth-period-distribution function, and no
mass-dependence of binary properties, can be formulated which
reproduces the empirical data.

\begin{figure}
\begin{center}
\rotatebox{0}{\resizebox{0.6 \textwidth}{!}
{\includegraphics{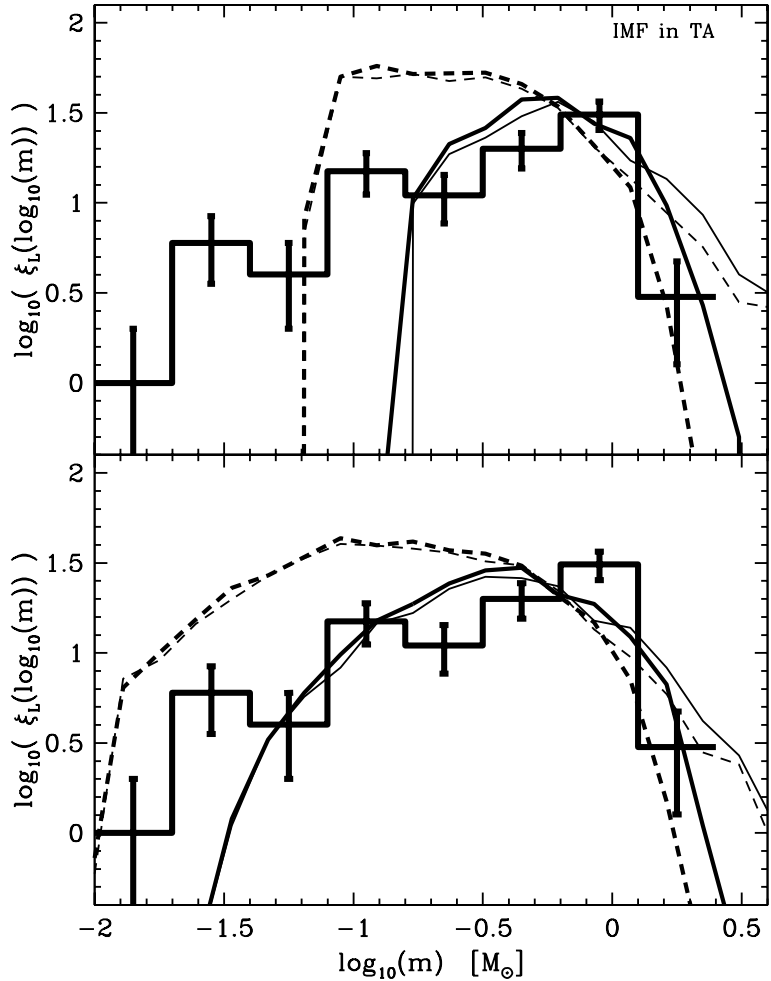}}}
\vskip -5mm
\caption
{The thick histogram shows the observed MF of stars and BDs in TA
arrived at by Luhman et al. (2003).  These data are compared to the
standard IMF (eq.~2 in KB1) without (upper panel) and with BDs (lower
panel), assuming all stars and BDs can be observed (dashed curves), or
only unresolved binary-system masses can be measured (solid
curves). Binary systems are constructed by random pairing from the IMF
in all cases, and the results are shown after pre-main sequence
eigenevolution (see \S~\ref{sec:ev}) has been allowed to act. The
models have been generated with 4000~stars and have been scaled to the
data using the same scale-factor in all cases.  {\bf Upper panel:} The
thick curves are for stellar masses in the mass range $0.07\le
m/M_\odot \le 1.5$, which is applicable to TA, while the thin curves
show the models if $0.07\le m/M_\odot \le 50$, which is applicable to
rich clusters.  Note that inclusion of stars more massive than
$1.5\,M_\odot$ has a negligible effect on the binary-star MF at low
masses.  In the SM BDs need to be thought of as an additional
(dynamically unimportant, Kroupa 2002) population such that the
overall theoretical MF agrees with the empirical one. In this case the
BDs do not participate in the pairing to binary systems with stellar
primaries initially, although some star--BD binaries will result
through dynamical capture and partner exchanges in the groups and
clusters in which most stars form.  {\bf Lower panel:} The same as the
upper panel, apart from the lower mass-limit being $0.01\,M_\odot$.
The discrepancy of the SMwBDs between the initial system MF (solid
curve) and the empirical data for log$_{10}(m/M_\odot) < -1.4$ could
be easily aleviated by slightly increasing the MF power-law index
$\alpha_0$ in the BD mass regime. The results shown in
Fig.~\ref{fig:r} suggest this not to be necessary however.  }
\label{fig:mftaur}
\end{center}
\end{figure}

Fig.~\ref{fig:mftaur} compares the observed MF in TA with the SM
(upper panel) and with the SMwBDs (lower panel).  Both models assume
the standard IMF which can be written as a three-component power-law
(eq.~2 in KB1), $\xi(m)\propto m^{-\alpha}$, where $\xi(m)\,dm$ is the
number of stars and BDs in the mass interval $m$ to $m+dm$, and
$\alpha_0=+0.3$ for $0.01 - 0.08\,M_\odot$, $\alpha_1=+1.3$ for $0.08
- 0.5\,M_\odot$ and $\alpha_2=+2.3$ for $m>0.5\,M_\odot$.  The figure
plots the ``logarithmic MF'', $\xi_{\rm L}({\rm log}_{10}m) = m\,{\rm
ln}(m)\,\xi(m)$, where $\xi_{\rm L}\,d{\rm log}_{10}m$ is the number
of stars/BDs or systems in the interval log$_{10}m$ to
log$_{10}m+d{\rm log}_{10}m$, $\xi_{\rm L} \propto m^{\Gamma}$ and
$\Gamma=1-\alpha$.  The figure shows that {\it the measured system MF
in TA is indistinguishable from the standard initial system MF and
thus perfectly normal}.

For the Orion Nebula cluster, Muench et al. (2002) measure
$\alpha=2.2$ for $m>0.6\,M_\odot$, $\alpha=1.2$ for $0.12 < m/M_\odot
\le 0.6$ and $\alpha=0.3$ for $0.025 \le m/M_\odot \le 0.12$. This
system MF is virtually identical to the standard IMF. Since it is the
measured MF of unresolved binary systems the underlying stellar IMF
will be somewhat steeper (larger $\alpha$) than the standard IMF
(i.e. containing relatively more low-mass stars,
e.g. Fig.~\ref{fig:mftaur} for models extending to $50\,M_\odot$).  In
addition, the MF has been measured within the inner regions of the ONC
which may differ from the global MF since mass segregation is well
pronounced in the cluster (Hillenbrand \& Carpenter 2000). The global
ONC IMF may thus be steeper still than the standard IMF. Apart from
this caveat due to mass segregation, the uncertainties in mass
estimates for stars and young-stellar-mass objects that are $\simless
1$~Myr old unfortunately preclude firm conclusions on differences or
similarities in the measured MFs. The usual approach taken to estimate
masses of such young stars is to compare their locations in the HRD
with pre-main sequence contraction tracks. These are calculated by
assuming that the stars begin fully convective and in hydrostatic
equilibrium, whereas the collapse and accretion invalidate this
assumption for objects younger than about 1~Myr (Wuchterl \&
Tscharnuter 2003).  Systematic errors that may vary depending on the
hydrodynamical history and thus entropy deposition history of the
hydrostatic core in a kernel may therefore affect mass estimates for
such young stars and BDs. Given these two caveats we conclude that,
pending further analysis, {\it the MFs of TA and ONC can be considered
as very similar if not identical}.

Concerning the about 100~Myr old Pleiades cluster, Moraux, Kroupa \&
Bouvier (2003) find excellent agreement with the observed MF and the
model system MF obtained from the cluster-formation computations of
KAH that assume the standard IMF. 

\begin{figure}
\begin{center}
\rotatebox{0}{\resizebox{0.6 \textwidth}{!}
{\includegraphics{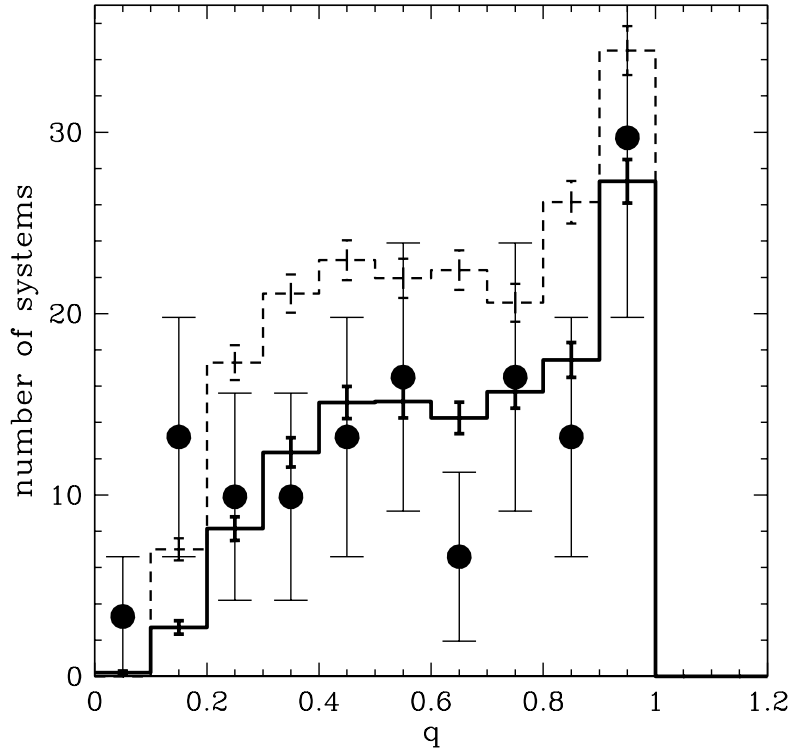}}}
\vskip -25mm
\caption
{The overall mass ratio distribution $q=m_s/m_p\le1$, where $0.1\le
m_p/M_\odot\le 1.1$ and $0.1\le m_s/M_\odot \le 1.1$ are the mass of
the primary and secondary, respectively.  The SM yields the dashed
histogram which is the initial mass-ratio distribution. It is
constructed by pairing both component masses at random from the
standard IMF (eq.~2 in KB1) and after taking into account pre-main
sequence eigenevolution (\S~\ref{sec:ev}).  Note the agreement with
the observational pre-main sequence constraints of Woitas et
al. (2001, their fig.~7).  After disintegration of the typical cluster
($N_{\rm bin}=200, R_{0.5}=0.8$~pc) found by Kroupa (1995) to
reproduce the Galactic-field population, the distribution evolves to
the thick solid histogram.  A small number of systems (about 10) with
$q=1$ result from eigenevolution.  The solid dots are observational
constraints, scaled to the model, from the 8~pc sample of Reid \&
Gizis (1997), with WD companions not counted for compatibility with
the model sample. These observational data are, however, likely to be
subject to a bias favouring brighter, $q\approx1$ systems, especially
so since their 8~pc sample is not complete (Henry et al. 1997).}
\label{fig:fq}
\end{center}
\end{figure}

\begin{figure}
\begin{center}
\rotatebox{0}{\resizebox{0.6 \textwidth}{!}
{\includegraphics{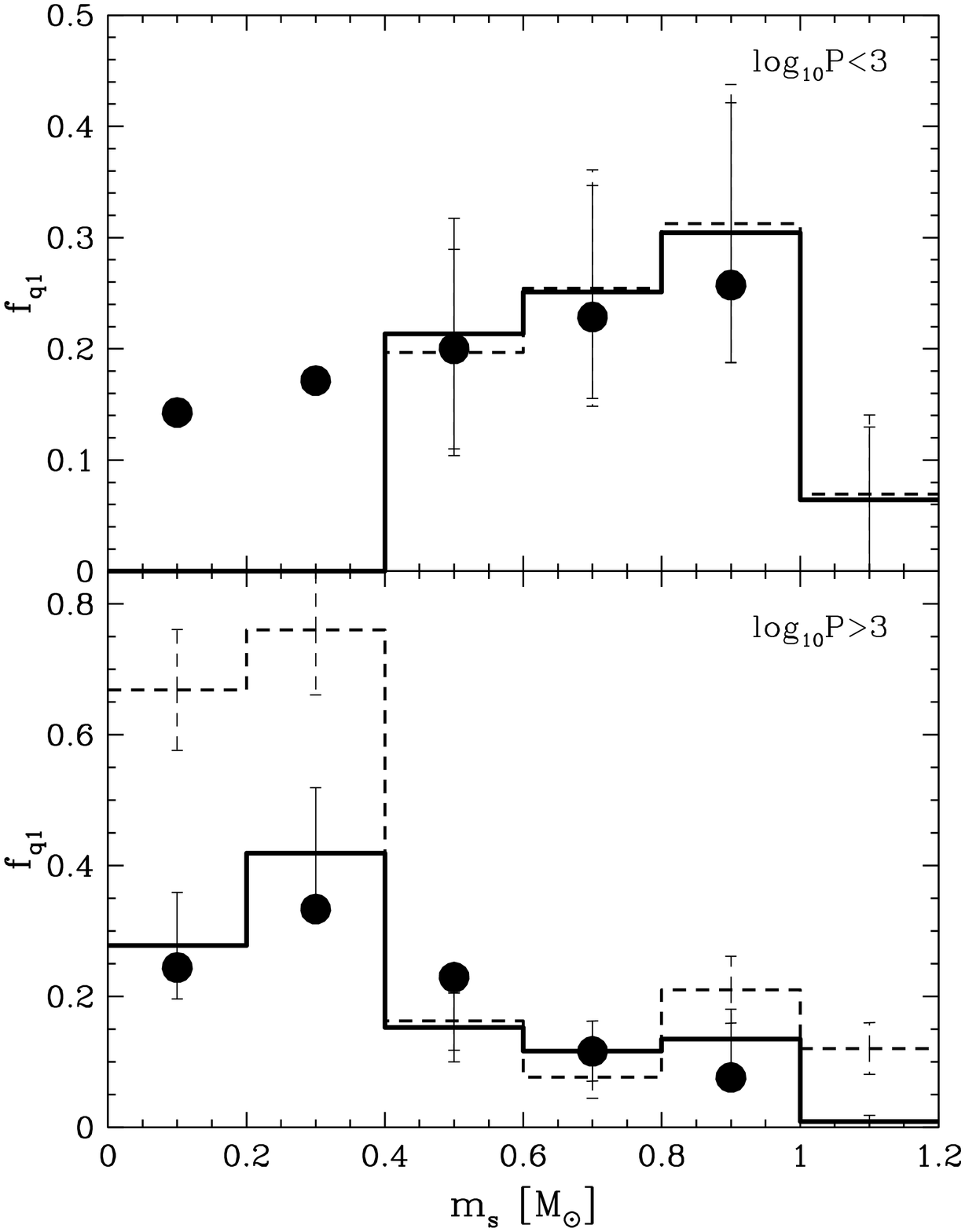}}}
\vskip 0mm
\caption
{The Galactic-field G~dwarf mass-ratio distribution, $q_1 = m_s/m_p$,
$0.9\le m_p/M_\odot \le 1.1$, $m_s \le 1.1\,M_\odot$.  {\bf Top
panel}: short period distribution of secondary star masses.  The
dashed histogram represents the initial distribution ($t=0$) after
pre-main sequence eigenevolution (\S~\ref{sec:ev}), and the solid
histogram is the final distribution.  It changes barely because the
stellar-dynamical interaction cross-section is too small for the
typical star-forming cluster.  The solid circles are G dwarf main
sequence short period binary star data (Mazeh et al. 1992). {\bf
Bottom panel}: the same as top panel but for long period systems for
which eigenevolution is insignificant. The solid dots are G~dwarf main
sequence long period binary star data (Duquennoy \& Mayor 1991). Note
that the primordial or birth distribution (before eigenevolution sets
in) in the upper panel is the same as in the lower
panel. Eigenevolution evolves the short-period mass-ratio distribution
to the form shown in the upper panel.}
\label{fig:fq1}
\end{center}
\end{figure}

Observations of low-mass, about 1~Myr old pre-main sequence stars in
low-density star-forming regions have established that most are in
binary systems (Duch\^ene 1999) with a flat mass-ratio distribution
for mass ratios $q\equiv m_s/m_p \simgreat 0.2$ (Woitas, Leinert \&
K\"ohler 2001), where $m_p$ and $m_s$ are the primary and secondary
mass, respectively. Random pairing from the standard stellar IMF also
gives such an approximately flat distribution (Fig.~\ref{fig:fq}).
Most stars appear to be born in modest clusters similar to the
``dominant mode cluster'' (\S~\ref{sec:intro}).  Disruption of binary
systems in such modest star clusters leads to fine agreement with the
overall mass-ratio distribution for Galactic-field binaries
(Fig.~\ref{fig:fq}), as well as with the mass-ratio distribution of
Galactic-field G-dwarfs (Fig.~\ref{fig:fq1}).

\begin{figure}
\begin{center}
\rotatebox{0}{\resizebox{0.6 \textwidth}{!}
{\includegraphics{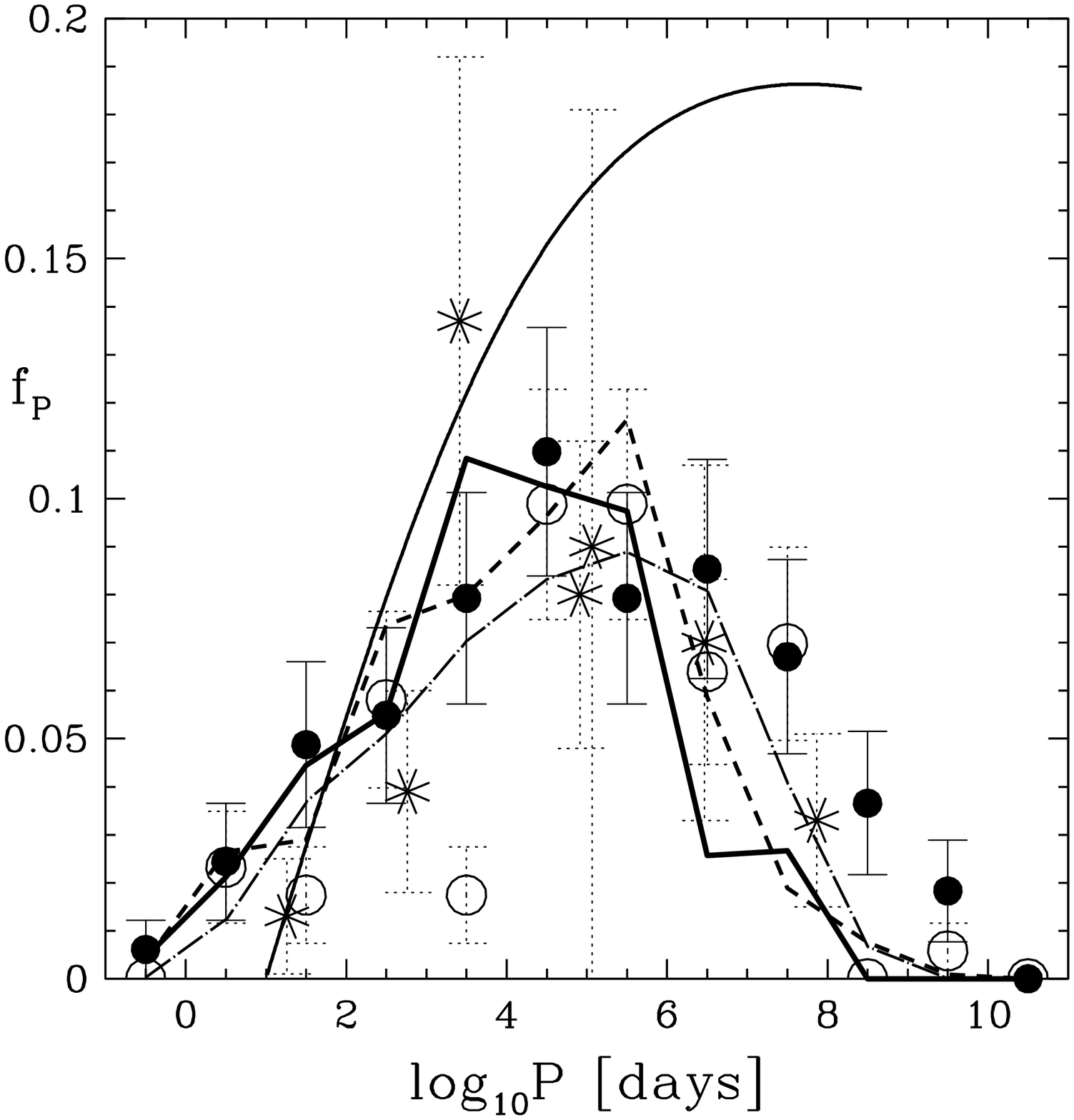}}}
\vskip -20mm
\caption
{Dependence of the period distribution of Galactic-field binaries on
primary mass.  Solid dots are data for G-dwarfs (Duquennoy \& Mayor
1991), open circles are for K-dwarfs (Mayor et al. 1992) and stars are
for M-dwarfs (Fischer \& Marcy 1992). The SM gives the solid
curve. This is the birth period distribution function (eq.~3 in KB1)
and is independent of primary mass. It fits the period-distribution
data of pre-main sequence stars (Fig.~\ref{fig:fp1} below). In modest
star clusters (and after eigenevolution) it evolves to the model
Galactic-field population shown by the solid line for G~dwarfs
($0.9<m_p/M_\odot \le1.1$), by the dashed line for K~dwarfs
($0.5<m_p/M_\odot\le0.9$) and by the dash-dotted line for M~dwarfs
($0.08<m_p/M_\odot\le0.5$).}
\label{fig:fp0}
\end{center}
\end{figure}

The orbital properties of late-type Galactic-field binaries do not
appear to vary with the mass of the primary. Thus, M-, K- and G-dwarfs
have indistinguishable period distribution functions
(Fig.~\ref{fig:fp0}). 
The distribution of
periods is given by
\begin{equation}
f_{\rm P} = {N_{\rm bin, P} \over N_{\rm sys}},
\label{eq:fp}
\end{equation}
where $N_{\rm sys} = N_{\rm bin}+N_{\rm sing}$ is the number of
systems with primaries in the corresponding mass range, while $N_{\rm
bin, P}$ is the number of binaries in the bin log$_{10}P$ (the period
$P$ is in days throughout this text) with primaries in the same mass
range.  The SM, which assumes that all primaries have a birth period
distribution function (eq.~3 in KB1) that is consistent with the
pre-main sequence data independently of the primary mass, matches the
1~Myr old pre-main sequence data and reproduces the M-, K-, and
G-dwarf period distributions observed for Galactic-field systems after
dissolution of the modest clusters.
The resulting model distribution of specific angular momenta of the
primordial binary population forms a natural extension to low values
of the observed specific angular momentum distribution of molecular
cloud cores (Kroupa 1995). We note that the binaries are constructed
by assigning each system a period, an eccentricity and mass ratio,
rather than binding energy, angular momentum and mass ratio. This is
done because the period-distribution functions can be readily derived
observationally.

\begin{figure}
\begin{center}
\rotatebox{0}{\resizebox{0.6 \textwidth}{!}
{\includegraphics{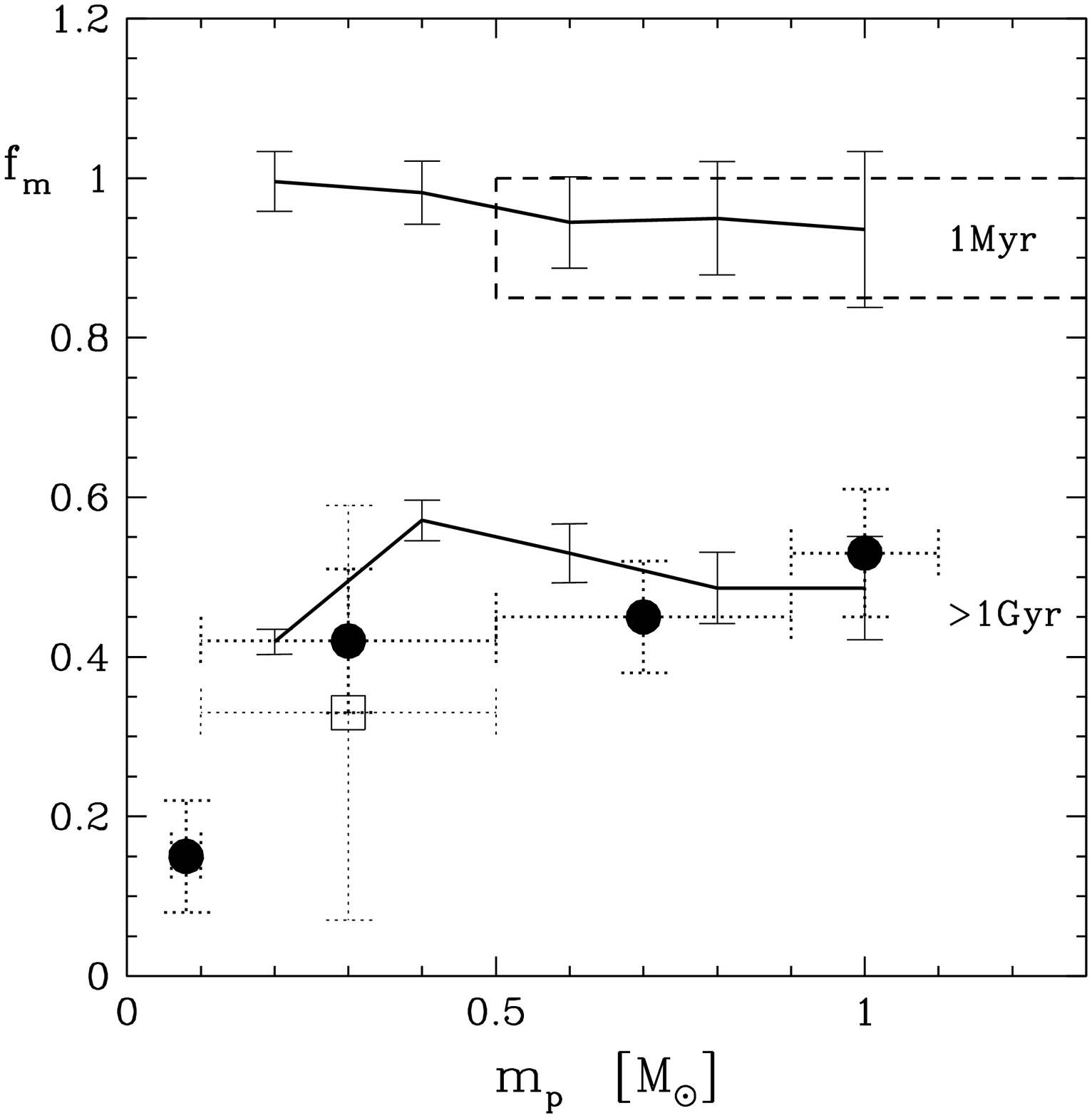}}}
\vskip -20mm
\caption
{Multiplicity fraction, $f_m$, versus primary mass, $m_p$, for the SM
compared with observational data (from Kroupa 1995). The upper solid
curve is the initial TA-like model, while the lower solid curve shows
the Galactic-field model population.  It evolves from the initial
population through stellar-dynamical processes in modest star
clusters.  About 1~Myr old pre-main sequence data are indicated by the
dashed rectangle (e.g. Duch\^ene 1999). These data are based on an
interpolation of the observed restricted period ranges
(Fig.~\ref{fig:fp1}).  The about 5~Gyr old Galactic-field population
is shown as solid circles: from right to left: G-dwarfs (Duquennoy \&
Mayor 1991), K-dwarfs (following Leinert et al. 1993), M-dwarfs
(Fisher \& Marcy 1992).  The estimate of the M-dwarf binary fraction
by Kroupa, Tout \& Gilmore (1993) is indicated by the open square.
The recent data for Galactic-field very-low-mass stars and massive BDs
is the $m_p\approx0.08\,M_\odot$ datum from Close et al. (2003) and
Gizis et al. (2003). It appears to violate the near-constant $f_m$ vs
$m_p$ relation, and is confirmed by the independent survey of Bouy et
al. (2003). }
\label{fig:fpm}
\end{center}
\end{figure}

The dependence of the binary fraction on the mass of the primary star,
$f_m$, is plotted in Fig.~\ref{fig:fpm}. There is a significant
difference between the Galactic-field binary population and the
pre-main sequence population, but the difference is accounted for very
well by the SM assuming most stars form in modest star
clusters. Particularly noteworthy is the probably significant
empirical change in $f_m$ near the hydrogen burning mass limit. This
change is complemented by the observation that M~dwarfs have a similar
period distribution as G~dwarfs (Fig.~\ref{fig:fp0}), while
very-low-mass stars and massive BDs have a period distribution
confined to log$_{10}P<4.9$ (\S~\ref{sec:dat}, $a<20$~AU and assuming
a system mass of $0.16\,M_\odot$). Note that we do not include the Reid
\& Gizis (1997) M~dwarf datum in Fig.~\ref{fig:fpm} because that
survey is incomplete (Henry et al. 1997). Incompleteness has the
effect that not all low-mass companions are seen leading to an
underestimate of the binary fraction. The detected binary systems can
be used to construct a mass-ratio distribution, but will be biased to
$q\approx1$ systems because these are brightest and thus more easily
seen. This may be one reason for the peak in the empirical data
evident in Fig.~\ref{fig:fq}. The SM reproduces this peak, but in this
case it is a result of pre-main sequence eigenevolution. Larger
samples will be needed to better constrain the overall mass-ratio
distribution.  We mention for completeness that SD03 include the Reid
\& Gizis (1997) datum in their fig.~2 thus giving the appearance of a
smoother variation of $f_m$ with $m_p$ than evident in
Fig.~\ref{fig:fpm}, but they do not match the pre-main sequence data.

The SM thus successfully describes the main sequence population in the
Galactic field, as well as the pre-main sequence population in TA but
also in the ONC when dynamical evolution is taken into account
(Kroupa 1995; Kroupa, Petr \& McCaughrean 1999, hereinafter KPM).
{\it The general result from this work so far is that the properties
of the stellar products of star-formation appear to be surprisingly
insensitive to the physical conditions of the cloud core in which the
population forms}.

The present formulation of the SM does not include triple and
higher-multiple stellar systems. The number of such multiples that
form in modest clusters through triple-star and binary--binary
encounters is too small to account for the observed number (Kroupa
1995). This is not a serious flaw however because higher-order
multiples make-up only about 30~per cent or less of all multiples, and
hierarchical quadruples are essentially two weakly bound binaries, but
for completeness sake future versions of the SM may need to
incorporate primordial triples. However, before extending the SM it
will be necessary to calculate the fraction of higher-order multiples
that are the remnants of dissolved clusters (de la Fuente Marcos 1998)
and TA-like aggregates using dynamical population synthesis. This
fraction will depend on the MF of star clusters, i.e. on how many
low-mass clusters are born per massive cluster in a population of
clusters.

\subsubsection{The SMwBDs (hypothesis~B)}
\label{sec:SMwBDs}

The recent detection of sub-stellar objects in a variety of
environments, however, prompts for the inclusion of BDs into the
model. The simplest scenario in concordance with the assertion that
BDs form like stars, is to simply extend the standard model to include
BDs. That is, the SMwBDs assumes that binaries are born by
fragmentation of a cloud kernel and that the two fragments have masses
sampled randomly from the IMF which extends into the BD mass range,
and that the primordial star--star, star--BD and BD--BD binaries have
the same period distribution function.  Note that this changes the
orbital distribution functions of star--star binaries since in the
SMwBDs some stellar primaries have BD rather than stellar companions.
The star--star binary fraction is thus lower in the SMwBDs than in the
SM.

\subsubsection{Eigenevolution}
\label{sec:ev}

The standard model, originally formulated without BDs, takes into
account that close (orbital periods $P\simless 10^3$~days) pre-main
sequence binaries evolve through system-internal processes (termed
``eigenevolution''). This is necessary to account for the observed
correlations between eccentricity, orbital period and mass ratio
(upper vs lower panel in Fig.~\ref{fig:fq1}).  The birth period
distribution function (eq.~3 in KB1 and the solid curve in
Fig.~\ref{fig:fp0}) evolves instantly in the model, but within a few
orbital times in reality, to the initial distribution, shown as the
dotted histograms in Fig.~\ref{fig:fp1} below.  While the resulting
changes to the IMF are negligible (Kroupa 2001), eigenevolution does
lead to some BDs in short-period systems acquiring stellar
masses. This leads to an overproduction of short-period star--star
binaries over the original formulation of the model, with the
consequence that the star--star period distribution function flattens
in the SMwBDs. This is evident in Fig.~\ref{fig:fp2} below.  This
problem can, in principle, be removed by not allowing short-period BD
companions to grow in mass during the formation phase. But such a
change to the model will only be attempted once its failures in other
respects have become more apparent.

\section{Stellar-dynamical models}
\label{sec:mods}

This section briefly describes the stellar-dynamical models of the TA
groups and of the ONC. We note that we only set-up $N$-body models of
the SMwBDs, because the SM follows trivially in the sense that the
period distribution functions remain very similar for star--star
systems in the SM as for the combined star--star and star--BD
distributions in the SMwBDs. This is the case because BD systems only
comprise a small fraction of the whole population in the SMwBDs
(table~2 in KAH). BDs play an unimportant role for the dynamical
evolution of clusters with normal IMFs comprising less than 5~per cent
in mass (Kroupa 2002), which is why BDs can be neglected altogether in
the SM where only the multiplicity properties of the stars are of
interest. BDs can be added, in the first instance, in the form of
Gedanken experiments, essentially by adding the required number of
test-particles with the required kinematical properties.

Six TA-like model aggregates (T0--T5) are constructed with
140~numerical renditions of each (table~1 in KB1).  The initial
conditions approximate the physical properties of the observed groups
in TA.  The aggregates contain 25 binaries initially and are embedded
in a time-varying gas potential.  The gas potential contains twice the
mass in stars in each case, and has the same density distribution as
the stars.

Aggregate (or cluster) T0 has an initial half-mass radius
$R_{0.5}=0.3$~pc while T1 has $R_{0.5}=0.8$~pc.  The gas potential is
removed through the accumulating outflows (Matzner \& McKee 2000)
after 0.5~Myr with an exponential time-scale of $\tau_{\rm M}=1$~Myr
in both models.  The crossing time through each aggregate is about
$t_{\rm cross}=1.2$~Myr (T0) and 5.2~Myr for T1.  Models T0 and T1
assume the standard IMF, $\alpha_0=+0.3$ for $0.01 - 0.08\,M_\odot$,
$\alpha_1=+1.3$ for $0.08 - 0.5\,M_\odot$ and $\alpha_2=+2.3$ for
$m>0.5\,M_\odot$.

Models T2--T5 have, like T0, an initial half-mass radius
$R_{0.5}=0.3$~pc, but their gas is removed after an embedded phase of
2~Myr with $\tau_{\rm M}=2$~Myr.  The IMF in models~T2--T5 is
identical to the standard IMF apart from having fewer BDs:
$\alpha_0=-4.2$ (T2), $\alpha_0=-3.0$ (T3), $\alpha_0=-1.5$ (T4) and
$\alpha_0=-0.5$ (T5).

Two models studied by KAH of cluster formation that fit the ONC and
the Pleiades are also used here. These have initially 5000~binaries
and $R_{0.5}=0.45$~pc (model~A, $t_{\rm cross}=0.13$~Myr) and
$R_{0.5}=0.21$~pc (model~B$, t_{\rm cross}=0.038$~Myr), and are again
initially embedded in a gas potential with twice the mass in stars and
with the same density profile as the stellar component. The gas is
removed due to the action of the O~stars on a thermal timescale
($<t_{\rm cross}$) after an embedded phase lasting 0.6~Myr. Both
models assume the SMwBDs with the standard IMF ($\alpha_0=+0.3$), so
changes in populations induced purely by the dynamical evolution can
be studied by a comparison of the TA and the ONC/Pleiades models. Wide
binaries are not able to form in such an environment where the tidal
field pulls the two fragments apart thus producing two single stars
(Horton, Bate \& Bonnell 2001). This is equivalent to what happens in
the standard model in which wide binaries are also immediately pulled
apart due to the cluster tidal field. 

Stellar-dynamical models of young clusters, such as constructed by KAH
and KB1, essentially assume that the entire initial population is
suddenly created in one go.  This is not correct for clusters that
contain a significant fraction of objects that are younger than the
crossing time, but for older cluster it is consistent with the very
rapid formation of clusters on a free-fall time scale and the rapid
dispersion of their embedding clouds (Elmegreen 2000; Hartmann,
Ballesteros-Paredes \& Bergin 2001; Hartmann 2003). In reality there
will be an age-spread comparable to the length of time required to
build-up the cluster until cessation of further star-formation
activity.  The age-spread is likely to be a significant fraction of
the age for very young objects such as TA and the ONC. Thus, the ONC
has an average age of about 1~Myr and an age-spread of at least a few
0.1~Myr (Hillenbrand 1997).  This can be accommodated in pure $N$-body
models by allowing new binaries to be added to the cluster following a
prescribed star-formation history.  However, this is not necessary,
because the crossing time through the embedded cluster is about
0.13~Myr or shorter implying that the vast majority of stars born in
the ONC will have experienced a few orbits through the cluster by the
time they are observed. Since binary destruction occurs on a
crossing-time scale (Kroupa 2000), the overall properties of the
binary population will therfore be essentially identical to a co-eval
population.  Sudden creation is thus a good approximation for a
cluster which contains a significant fraction of objects that are
older than a few crossing times.  The models as set up by KAH
therefore reproduce the properties of the stellar population very
well. The TA models do not evolve significantly dynamically (KB1) so
that the binary properties mostly remain age invariant.

The evolution is integrated for 40~Myr for the TA-like aggregates and
for 150~Myr for the ONC/Pleiades models with the high-precision {\sc
Nbody6}-variant {\sc GasEx} (KAH) which allows accurate treatment of
close encounters and multiple stellar systems in clusters through
special mathematical transformation techniques of the equations of
motion (regularisation), so that a force-softening parameter is not
needed. Subsequently to the $N$-body integration, a data-reduction
software package is used to distill the data presented in this
contribution.

The evolution of the aggregates is described in KB1. Briefly, the
aggregates largely dissolve within 10~Myr, thereafter only a
long-lived group containing a few systems ($=$ binaries plus single
stars) remains within the central 1~pc region. Binary destruction is
inefficient, and in all cases the binary proportion remains
significantly higher than is observed in the Galactic field. This
means that TA-like star-formation events did not contribute
significantly to the Galactic-field population.

\begin{figure}
\begin{center}
\rotatebox{0}{\resizebox{0.6 \textwidth}{!}
{\includegraphics{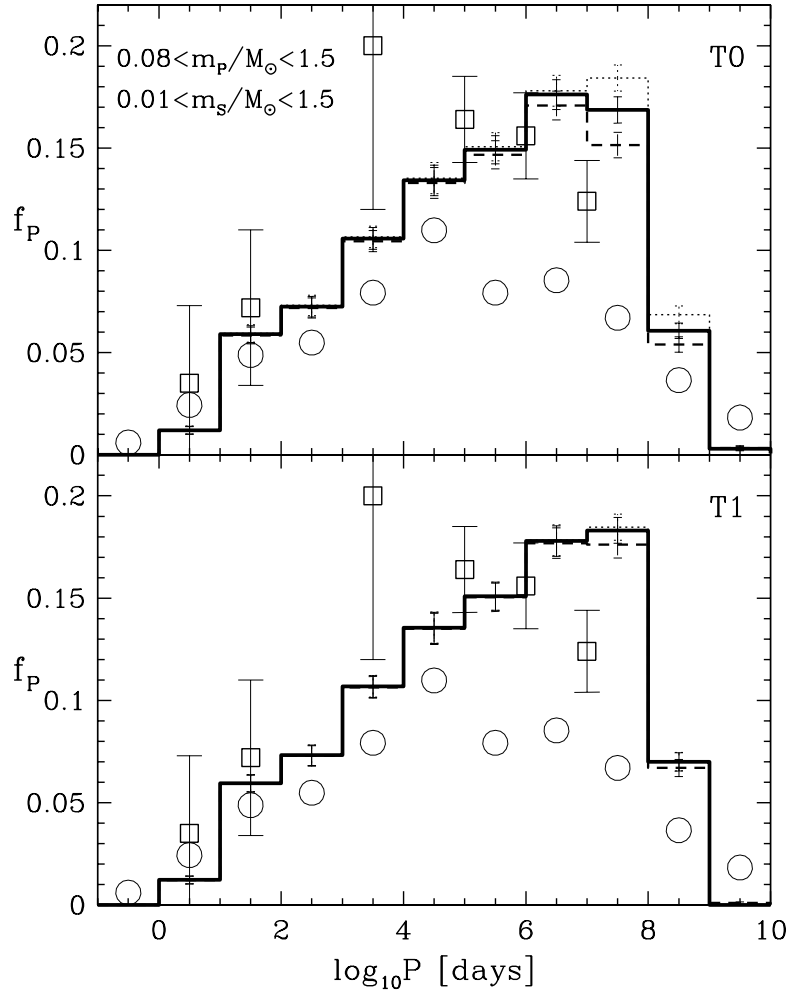}}}
\vskip 0mm
\caption
{The distribution of periods for late-type stellar primaries in
TA-like aggregates according to the standard model.  The thin dotted
histograms are the initial distributions (after pre-main sequence
eigenevolution (\S~\ref{sec:ev}) and derived from the solid curve in
Fig.~\ref{fig:fp0}), the thick solid histograms are for
$t\approx1$~Myr, while the dashed histograms are for $t\approx 15$~Myr
(the final distributions). The histograms are averages from
140~renditions per model, the errorbars being standard deviations of
the mean. Model~T0 has an initial half-mass radius $R_{0.5}=0.3$~pc
while model~T1 has $R_{0.5}=0.8$~pc.  T1 shows negligible evolution.
The faint open circles are observational data for Galactic-field
G~dwarf primaries (Duquennoy \& Mayor 1991), while the open squares
are pre-main sequence data from TA and similar environments (Mathieu
1994; Richichi et al. 1994; K\"ohler \& Leinert 1998). Note that the
initial distribution is in very good agreement with the pre-main
sequence data. This figure is to be compared with figs.~10 and 11 in
KAH that show the corresponding distributions in ONC and Pleiades-like
clusters.  Note also that here $N_{\rm bin, P}$ includes BD
companions, but the same initial distribution is obtained for
star--star binaries in the SM (Kroupa 1995 and KPM).  }
\label{fig:fp1}
\end{center}
\end{figure}

The initial period distribution function that enters all models
considered here is shown as the dotted histogram in
Fig.~\ref{fig:fp1}.  The pre-main sequence data appear to show a
different trend ($f_{\rm P}$ decreasing with increasing log$_{10}P$
for log$_{10}P>3$), but close inspection shows that only two data
points deviate from the initial model and that this deviation is at a
level of less than two-sigma.  The figure also shows that the
distribution does not evolve significantly in TA-like aggregates.
The distribution does evolve significantly in the ONC-like models~A
and~B. By 1~Myr the evolution has created period distribution
functions of late-type stellar primaries that agree with the observed
distributions in the ONC and the Pleiades (figs.~10 and~11 in KAH).

The BD velocity distribution function resulting from the TA models has
a tail of high-velocity ($1\simless v \simless 10$~km/s) BDs, which
are nearly exclusively single and comprise about 15~per cent of all BD
systems. The high-velocity tail results from ejections of BDs from
short-lived three- and four-body systems that form through binary
encounters in the TA-like groups.  In this respect, the SMwBDs cannot
be distinguished from the embryo-ejection model. The embryo-ejection
model, however implies the majority of BDs to be single and to have
velocities $v\simgreat 1$~km/s.  In sharp contrast, for TA the SMwBDs
implies that the slow-moving BD systems with a velocity $v\simless
0.5$~km/s, which amount to about 60~per cent of all BD systems, retain
a high binary proportion of 60~per cent or larger.

\section{The number of brown dwarfs per star}
\label{sec:noBDs}

The first question which we address here within the framework of
hypothesis~B is if stellar-dynamical evolution of TA-like aggregates
can explain the Briceno et al. (2002) result without calling for a
different IMF.  The expectation is that this may be the case, because
binary--binary encounters in the ONC will have been much more
destructive due to the significantly higher stellar density and
shorter crossing time than in the TA aggregates.  Indeed, the ONC is
known to have a significantly smaller binary proportion than the TA
population (Prosser et al. 1994; see also KAH).  An observer would
thus see more BDs per star in the ONC than in the TA aggregates simply
because the BD companions have been freed from their stellar primaries
and because star--BD and BD--BD binaries have been disrupted
preferentially owing to their weaker binding energy.  This issue of
apparent (but not true) IMF variations in clusters has been much
stressed elsewhere (Kroupa 2001), but it is important to return to
this notion in a case-by case study.

To investigate the issue of an apparent depletion of BDs in TA
relative to the ONC population, the ratio
\begin{equation}
{\cal R}_{\rm obs} = {N_{\rm sys}(0.02-0.08\,M_\odot) \over 
           N_{\rm sys}(0.15-1.0\,M_\odot)}
\label{eq:r}
\end{equation}
is computed for each of the models; $N_{\rm sys}$ is the number of
systems with primaries in the respective mass range.  The lower mass
limit, $m=0.02\,M_\odot$, is given by the observational limits of the
Briceno et al. (2002) survey.  This ratio is similar to the two ratia
considered by Briceno et al. and has the advantage of not being
sensitive to detailed structure in the IMF, as stressed by Briceno et
al.  The ratio ${\cal R}_{\rm obs}$ used here does not include stars
more massive than $1\,M_\odot$, because star formation in TA is biased
against the production of such stars given the limited supply of gas.
The data provided by Briceno et al. imply for TA,
\begin{equation}
{\cal R}_{\rm obs,TA} = {10 \over 59} = 0.17\pm0.06,
\label{eq:rta}
\end{equation}
while for the central part of the ONC,
\begin{equation}
{\cal R}_{\rm obs,ONC} = {47 \over 125} = 0.38\pm0.06.
\label{eq:ronc}
\end{equation}

\begin{figure}
\begin{center}
\rotatebox{0}{\resizebox{0.7 \textwidth}{!}
{\includegraphics{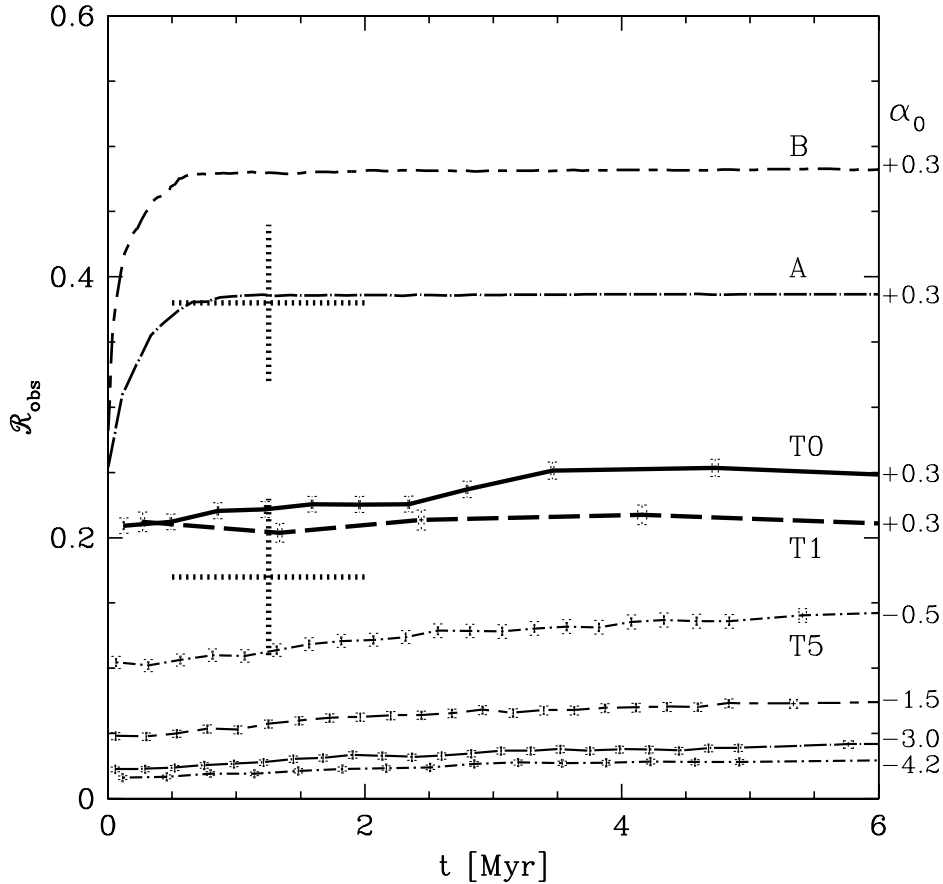}}}
\vskip -30mm
\caption
{The evolution of the ratio ${\cal R}_{\rm obs}$ ($=$ the number of BD
systems per late-type stellar system) for the various models as
indicated in the figure (the unlabelled curves are from bottom to top:
models~T2,~T3 and~T4). Each curve (except A\&B) is the average of
140~random number renditions per model, and the errorbars are standard
deviation of the mean values.  The upper dashed cross is ${\cal
R}_{\rm obs,ONC}$ (eq.~\ref{eq:ronc}) for the central part of the ONC,
while the lower dashed cross is the Briceno et al. (2002) result
${\cal R}_{\rm obs,TA}$ (eq.~\ref{eq:rta}) for TA. The uncertainty in
age is taken to be 1.5~Myr for both.  Note that models T0, T1, A and B
have the same IMF, and in all cases the SMwBDs has been used.}
\label{fig:r}
\end{center}
\end{figure}

Fig.~\ref{fig:r} plots the evolution of ${\cal R}_{\rm obs}$ for each
of the KB1 models, as well as the two models calculated by KAH that
reproduce the ONC and the older Pleiades.  The figure demonstrates
that the disruption of binary systems may lead to the observed
apparent variation of the relative number of BD systems: identical
initial stellar and BD populations lead to very different values of
${\cal R}_{\rm obs}$, depending on the degree of dynamical
evolution. Thus, models~A and~B reproduce ${\cal R}_{\rm
obs,ONC}$. The initial rapid increase of ${\cal R}_{\rm obs}$ is due
to the disruption of binary systems on a crossing time-scale (fig.~9
in KAH).  It is interesting that model~A yields a somewhat better fit,
as this model also reproduces the radial density profile and the
binary period distribution function observed in both the ONC and the
Pleiades better than model~B.  The two models~T0 and~T1 of TA-like
aggregates reproduce the Briceno et al. (2002) TA datum very well. The
much smaller ${\cal R}_{\rm obs}$ is a result of the smaller fraction
of disrupted BD systems in these models.  The figure additionally
plots the results from the other TA-like models (T2--T5) that assume
non-standard IMFs in the BD regime (fewer BDs), and we note that
model~T5 ($\alpha_0=-0.5$) is also consistent with the data. IMFs in
the BD regime with too steep a slope ($\alpha_0<-1.5$) can be rejected
with high confidence.

As suggested above, an alternative interpretation of the Briceno et
al. result is thus that the birth population in the ONC was in fact
identical to that in TA, but that stellar-dynamical encounters
destroyed a large number of primordial binaries in the ONC leading to
the freeing of BDs. An observer would see different numbers of BDs per
star in both environments.  {\it The discovery of a significantly
lower frequency of BDs in TA does therefore not, by itself, constitute
evidence for a variable IMF}.  Additional diagnostics are needed to
infer a difference between the BD population in TA and in the ONC.

\section{Binary properties}
\label{sec:binprop}

Although the observed relative number of BDs and stars does not imply
a non-universal IMF, we now investigate the binary properties
(especially the presence of BDs in wide systems) predicted by our
models and compare them to observed properties. We here focus on
the distribution function of orbital periods and semi-major axes for
late-type stars and BDs.  

\subsection{A brief summary of available observational constraints} 
\label{sec:dat}

The observational pre-main sequence data plotted as open squares in
Fig.~\ref{fig:fp1} only include star--star binaries.  

Star--BD binaries are extremely rare in TA at separations of about~150
to~1000~AU (White \& Ghez 2001), and only one has been found
(GG~Tau~Bb, White et al. 1999). Using HST spectroscopy Hartigan \&
Kenyon (2002) find no BD companions to stars with separations $s=15$
to~150~AU among 20~systems in TA. However, this study relies on
relatively bright (i.e. massive) companions to obtain spectra, which
could impose a bias such that their sample is probably not complete
and may therefore not indicate a true absence of BD companions.

While BDs are very rarely companions to stellar primaries with
separations less than a few AU (the so-called ``brown dwarf desert``,
Marcy \& Butler 2000; Halbwachs et al. 2000), the frequency of wide
star--BD systems for field and open cluster stars is not as clear:
Reid \& Gizis (1997) found no BD companion in the separation range
5--200~AU to stars in the Hyades while Gizis et al. (2001) suggested
that BDs are quite commonly found as wide ($\simgreat 1000$\,AU)
companions to nearby field stars. Both studies, however, are limited
by large statistical uncertainties. 

The highly sensitive adaptive optics study of~39 very-low-mass
M8.0--L0.5 solar-vicinity dwarfs by Close et al. (2003) reveals 9
companions and a sensitivity corrected binary fraction of $15\pm7$~per
cent with mass ratios $q>0.7$ although $q=0.5$ systems should have
been detected.  The semi-major axis ($a$) distribution is narrow with
a peak near 4~AU; orbits with $a>20$~AU are not present. These
findings are in excellent agreement with the similar surveys made by
Gizis et al. (2003) and Bouy et al. (2003), and with the survey of the
Pleiades cluster by Martin et al. (2000, 2003). These results stand in
contrast to the binary fraction of slightly more massive
M0--M4~dwarfs, $32\pm9$~per cent (Fischer \& Marcy 1992) that also
show a broad semi-major axis range with a maximum near 30~AU, very
similar to G~dwarfs.  HST imaging of 10~BDs selected from a
magnitude-limited search of the 2MASS data base lead Burgasser et
al. (2003) to detect two BD--BD binaries with $s=3.2$ and~1~AU, but no
companions with $s>10$~AU.  Most binary surveys have been quite
sensitive to wide separations, making any deficit of such wide systems
a real effect and not a mere observational bias. This applies
especially to our knowledge of TA for systems with separations larger
than approximately 100~AU, as well as to the surveys of Burgasser et
al. (2003), Close et al. (2003), Gizis et al. (2003), Bouy et
al. (2003) and Martin et al. (2000, 2003).

These data thus suggest a marked change of the binary properties near
the hydrogen-burning mass limit, as is also stressed by Close et
al. (2003), as is evident from Fig.~\ref{fig:fpm}. The origin of this
behaviour may be primordial or dynamically induced, which we address
in the following.

\subsection{Models of the ONC and Pleiades with the standard IMF}
\label{sec:onc}

\begin{figure}
\begin{center}
\rotatebox{0}{\resizebox{0.6 \textwidth}{!}
{\includegraphics{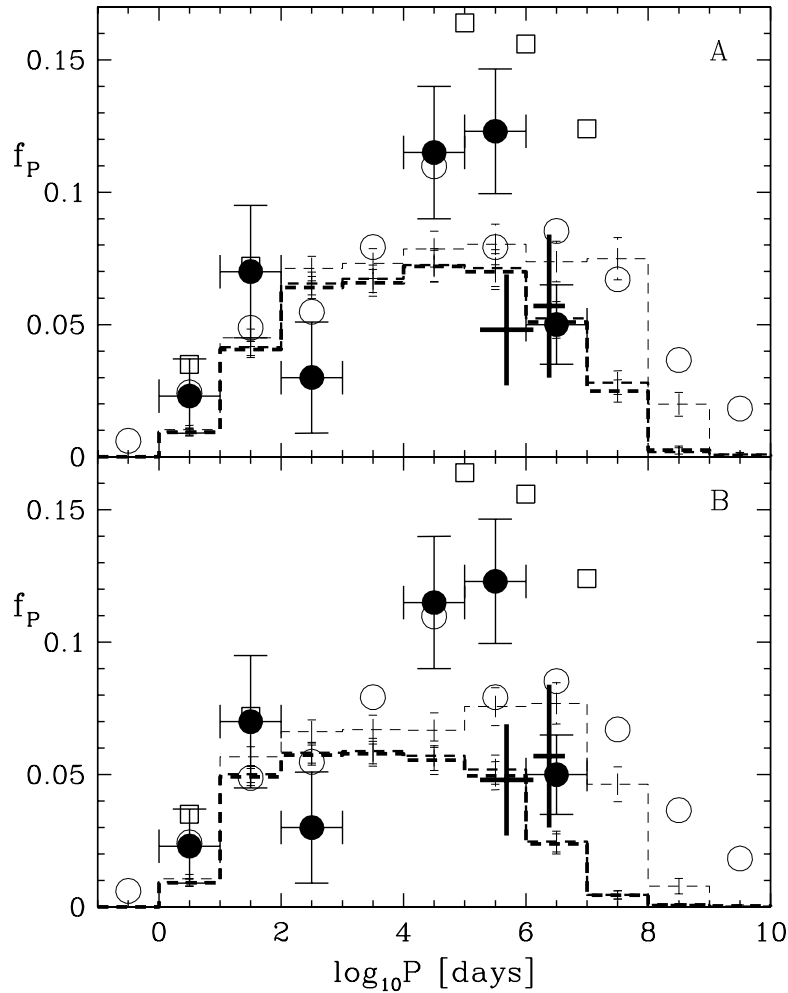}}}
\vskip 0mm
\caption
{The period distribution function (eq.~\ref{eq:fp}) of late-type
primaries ($0.08\le m_p/M_\odot\le 1.5$) with stellar secondaries
($0.08\le m_s/M_\odot\le 1.5$) initially (thin histograms), at 0.9~Myr
(medium histograms) and at 100~Myr (thick histograms). The models
include all stars within a radius of 15~pc from the centre of the
cluster, and the errorbars represent Poisson uncertainties.  The thick
solid crosses are observational data for the about 1~Myr old ONC (Petr
1998).  The solid circles are observational data for the Pleiades
cluster which has an age of about 100~Myr (Mermilliod et al. 1992;
Bouvier, Rigaud \& Nadeau 1997). The open circles are the
Galactic-field G~dwarf data and the open squares are pre-main sequence
data, mostly from Taurus--Auriga, as in Fig.~\ref{fig:fp1}.  }
\label{fig:fpAB}
\end{center}
\end{figure}

The SMwBDs with the standard IMF is in good agreement with
observational data available for the ONC and the Pleiades: There is
very good agreement with the number of BDs per star seen in the
central region of the ONC (Fig.~\ref{fig:r}), and the observed MF in
TA and the ONC (\S~\ref{sec:SM}), the Pleiades (Moraux et al. 2003)
and the Galactic field are all well consistent with the SMwBDs.  The
period distribution for star--star binaries is also in good agreement
with the data (Fig.~\ref{fig:fpAB}) for both clusters and both models
(although the initially less concentrated model~A appears to fit
somewhat better).  The finding is thus that {\it the SMwBDs with the
standard IMF is consistent with the available data for the ONC and the
Pleiades}.

However, the SM for the ONC and Pleiades also leads to agreement with
the period-distribution constraints. In the SM the initial period
distributions would lie above the initial models shown in
Fig.~\ref{fig:fpAB}, as is evident in figs.~10 and 11 in KAH. These
figures (10 and~11) are for the SMwBDs but would look like and evolve
as models that contain no BDs apart from a slightly reduced disruption
efficiency (i.e. slightly less evolution) due to the higher binding
energy of the stellar binaries. This is evident by comparing fig.~10
in KAH (with BDs) with the upper panel of fig.~4 in KPM (without BDs).
Additional $N$-body computations are thus not needed to explicitly
construct SMs. The evolution of the star--star binaries in the SM
leads to period distributions that match the observational data, as
shown in figs~10 and 11 in KAH.  

Thus, {\it a distinction between the SM and the SMwBDs cannot be made
yet on the basis of the available observational star--star binary data
in the ONC and the Pleiades}.  We would need BD data to achieve this.

\begin{figure}
\begin{center}
\rotatebox{0}{\resizebox{0.6 \textwidth}{!}
{\includegraphics{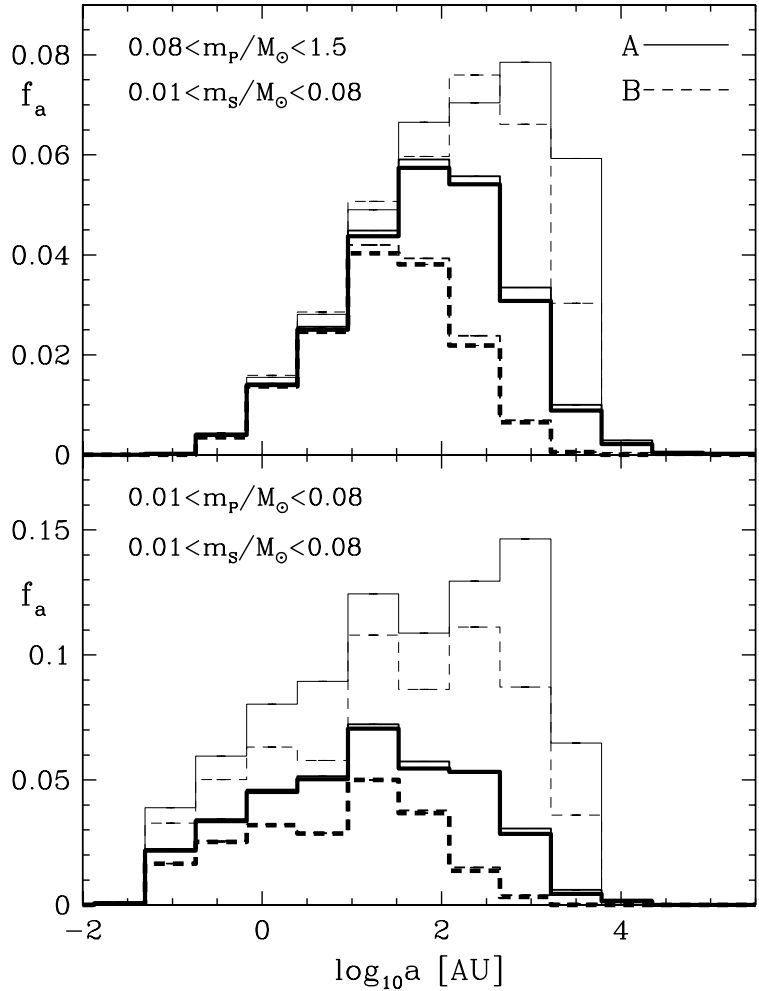}}}
\vskip 0mm
\caption
{{\bf Upper panel:} The semi-major axis distribution of star--BD
binaries in models~A and~B of the ONC and the Pleiades. {\bf Lower
panel:} The semi-major axis distribution of BD--BD binaries in both
models.  In both panels thin histograms are initial distributions,
medium histograms are at 0.9~Myr while thick histograms show the
100~Myr old distributions. All stars within 15~pc of the cluster
centre are counted.
}
\label{fig:fpAB_BDs}
\end{center}
\end{figure}

It may be possible to discard the SMwBDs if it can be shown that the
star--BD and the BD--BD semi-major axis distributions differ from
observational constraints in clusters.  The model predictions for
future observational tests are shown in Fig.~\ref{fig:fpAB_BDs}. For
instance 7--15~per cent of all low-mass stars in clusters should have
a BD companion with semi-major axis $a\simgreat 30$~AU.  
The overall binary fraction of BDs in Pleiades-like clusters is
approximately 20~per cent according to the SMwBDs. Using the location
of data points in the colour--magnitude diagram to identified binary
candidates, Pinfield et al. (2003) estimate the BD binary fraction in
the Pleiades to be about $50\pm11$~per cent which is somewhat higher
than but still consistent with the SMwBDs. Martin et al. (2000, 2003)
find a BD binary fraction of about 15~per cent and a lack of BD
binaries with $a>27$~AU in the Pleiades.  It will be necessary to
verify cluster membership and the nature of the suggested Pleiades BD
binaries and to measure the component separations as well as mass
ratios to allow firmer tests of the SMwBDs.

The possible success of the SMwBDs in the ONC and the Pleiades would
definitely be mired if a significant fraction of Galactic-field stars
derive from ONC- and Pleiades-like clusters.  The frequency of wide BD
companions to field stars (\S~\ref{sec:dat}) does not match the
predictions of our SMwBD models if stars are predominantly formed in
ONC-like clusters or in more modest clusters as is currently believed
(\S~\ref{sec:intro}).  In this case the problem would be even worse,
since the less dense modest clusters would lead to an even higher
surviving fraction of BD- and very-low-mass stellar binaries extending
to even larger separations than shown in Fig.~\ref{fig:fpAB_BDs}.
Only if most field stars formed in much denser clusters or have
suffered additional dynamical evolution can the SMwBD can be
reconciled with our current knowledge.

As an aside we note that Fig.~\ref{fig:fpAB} also demonstrates that
most of the evolution of the period distribution function occurs
during the embedded cluster phase prior to 1~Myr, being consistent
with observational data for a number of young populations.  For
example, the 2-3~Myr old cluster IC~348 already shows similar binary
properties as are found for Galactic-field main sequence dwarf stars
(Duch\^ene, Bouvier \& Simon 1999). The observed MF in IC~348 should
reflect this and be steeper by having a larger $\alpha$ than the
observed MF in TA (\S~\ref{sec:intro}, cf. Luhman et al. 2003c).
Indeed, Muench et al. (2003) find a nearly exact match of the IC~348
MF with that measured in the ONC, again suggesting a remarkable degree
of uniformity of the stellar populations. 

In conclusion we emphasise that the SM is consistent with the data in
the ONC and the Pleiades cluster, while wide star--BD and BD--BD
binaries appear to lead to inconsistencies of the SMwBDs with
observational data.  Since we already know that the SM also reproduces
the properties of stellar systems in the Galactic field we may be
encouraged to conclude that {\it the standard model without BDs (SM),
rather than the SMwBDs, may be the preferred description of initial
populations}. This possible conclusion depends in part on the
performance of the SMwBDs in TA, which is studied in the next section.

\subsection{Models of TA with the standard IMF}
\label{sec:T0T1}

\begin{figure}
\begin{center}
\rotatebox{0}{\resizebox{0.6 \textwidth}{!}
{\includegraphics{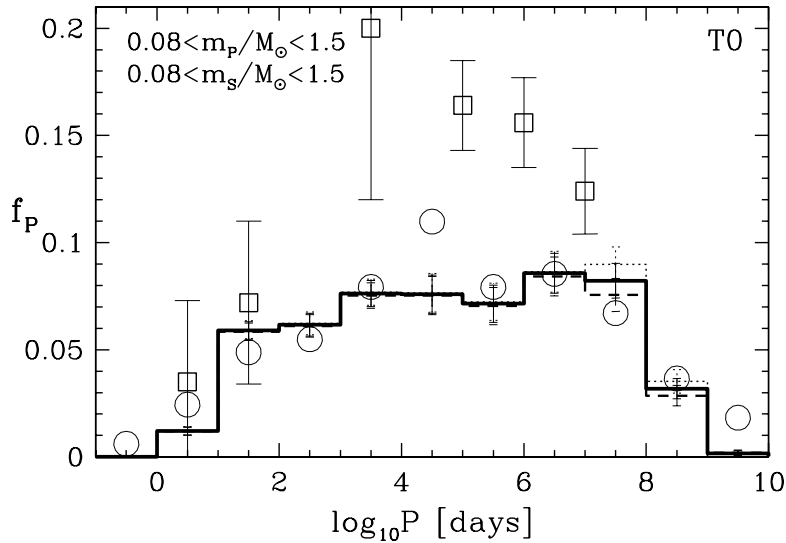}}}
\vskip -60mm
\caption
{As Fig.~\ref{fig:fp1} but showing the distribution of periods for
star--star binaries in TA-like aggregate T0 (T1 shows negligible
evolution). The pre-main sequence model does not agree with the
pre-main sequence data.  }
\label{fig:fp2}
\end{center}
\end{figure}

Fig.~\ref{fig:fp2} plots the initial and evolved period distributions
for star--star binaries in the TA-like aggregates.  The initial and
the later distributions do not fit the pre-main sequence data. This
results from the problem already alluded to in \S~\ref{sec:ev}, namely
that many stellar primaries now have BD companions as a result of
extending the standard model into the BD mass range. This reduces the
proportion of star--star binaries to the level evident in the figure,
because the period distribution function (eq.~\ref{eq:fp}), is
normalised to the total number of systems, $N_{\rm sys}$, with
primaries in the corresponding mass range ($0.08\le m_p/M_\odot \le
1.5$), while $N_{\rm bin, P}$ is the number of star--star binaries in
the bin log$_{10}P$ (star--BD systems are counted here as single
stars).

Furthermore, in the SMwBDs the shape of the initial star--star period
distribution function, plotted in Fig.~\ref{fig:fp2} (dotted
histogram), differs from the initial distribution plotted in
Fig.~\ref{fig:fp1}, by being flatter. This is a result of some BDs in
short-period binaries acquiring stellar masses due to the
eigenevolution model (\S~\ref{sec:ev}).

\begin{figure}
\begin{center}
\rotatebox{0}{\resizebox{0.6 \textwidth}{!}
{\includegraphics{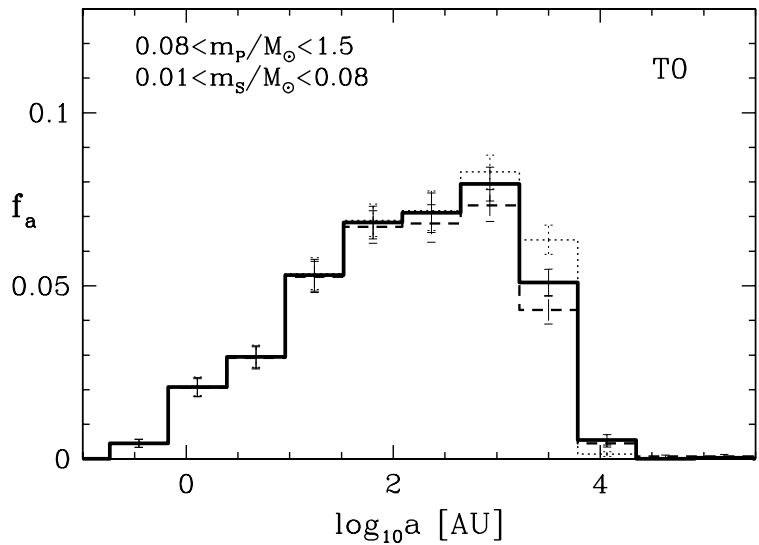}}}
\vskip -60mm
\caption
{The distribution of semi-major axes of BD companions to stellar
primaries according to the SMwBDs for TA-like aggregate T0 (T1 shows
negligible evolution).  The thin dotted histogram is the initial
distribution, while the thick solid histogram shows the distributions
at 1~Myr and the dashed histogram depicts the final ($\approx 15$~Myr
old) distribution. All histograms are averages of 140~model
renditions, and the error bars are standard deviation of the mean
values.}
\label{fig:semi1}
\end{center}
\end{figure}

The SMwBDs faces the more severe problem of leading to too many BD
companions to stellar primaries as a result of random pairing from the
IMF. The distribution of star--BD semi-major axes, plotted in
Fig.~\ref{fig:semi1}, implies that about 23~per cent of all late-type
stars in TA ought to have BD companions with separations in the range
$10^2-10^4$~AU. This means that about 13 of the 60~stars in the
Briceno et al. (2002) survey ought to have such companions, whereas
only GG~Tau~Bb is known (\S~\ref{sec:dat}). {\it The SMwBDs and the
standard IMF can thus be excluded with high confidence using this
argument}.

\begin{figure}
\begin{center}
\rotatebox{0}{\resizebox{0.6 \textwidth}{!}
{\includegraphics{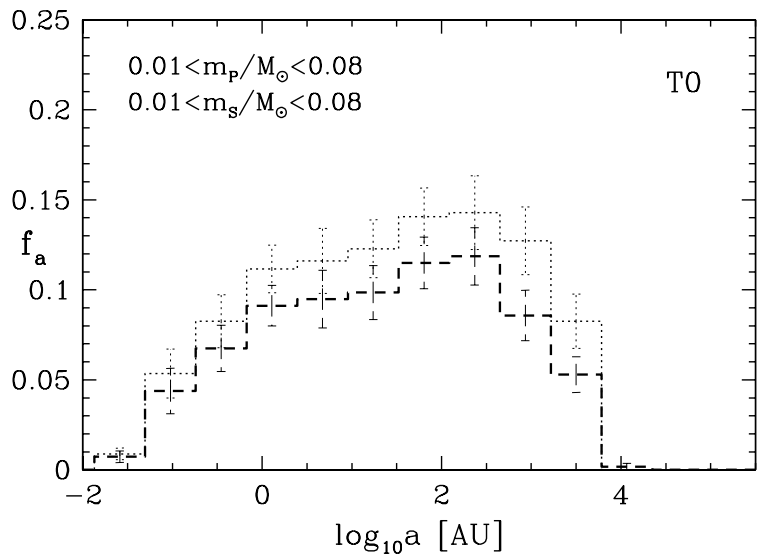}}}
\vskip -60mm
\caption
{The distribution of semi-major axes of BD--BD binaries according to
the SMwBDs for TA-like aggregate T0 (T1 shows negligible evolution).
The thin dotted histogram is the initial distribution, while the
dashed histogram depicts the final ($\approx 15$~Myr old)
distribution. All histograms are averages of 140~model renditions, and
the error bars are standard deviation of the mean values.}
\label{fig:semi2}
\end{center}
\end{figure}

As a final test of the SMwBDs, the semi-major axis distribution of
BD--BD binaries is plotted in Fig.~\ref{fig:semi2}: In TA about 30~per
cent of all BDs ought to have BD companions in the separation range
between~100 and $10^4$~AU. Given that 10 BDs have been found in TA
(Briceno et al. 2002) this implies that 3 companions ought to have
been seen. None have been found, but the statistical uncertainty is
too high to exclude the model with confidence using this particular
argument.

There are thus three tests of the standard model extended to include
BDs with the standard IMF (properties of the star--star, star--BD and
BD--BD binaries). In TA, at least two of these lead to arguments
against the model with a high statistical weight. The conclusion
therefore must be that {\it BDs are not paired with stars at random
from the standard IMF in TA}.

\subsection{A different IMF for BDs in TA ?} 
\label{sec:T5}

The simplest change to the model is to assume that the SMwBDs holds in
TA (i.e. stars and BDs are paired at random from the IMF and there is
no dependency of the orbital parameters on primary mass), and thus
remain in-line with the conclusions of Briceno et al. (2002) and White
\& Basri (2003) that the BDs in TA appear to have formed as the stars
did, but that the IMF in TA differs from the standard IMF.  The
standard IMF has $\alpha_0=+0.3, 0.01-0.08\,M_\odot$.  From
Fig.~\ref{fig:r} it is seen that model~T5 which has $\alpha_0=-0.5$,
is still in agreement with the Briceno et al. (2002) datum for the
number of BDs per star in TA. It also fits the ONC datum (by analogy
with models~A and~B).  Values $\alpha_0\simless -1.5$ lead to too few
BDs per star and are thus excluded.

It is therefore worthwhile to investigate if random pairing from the
IMF with $\alpha_0=-0.5$ can be brought into agreement with the
available data for TA. Such an IMF implies fewer BD companions to
stars, so that the star--star period distribution function will not be
as suppressed. The initial period distribution of all binaries is
identical to that shown in Fig.\ref{fig:fp1}, and the evolution is
also indistinguishable. That is, the disruption of binaries is very
inefficient in model~T5, just as it is in models~T0 and~T1, despite
the longer embedded phase.

\begin{figure}
\begin{center}
\rotatebox{0}{\resizebox{0.6 \textwidth}{!}
{\includegraphics{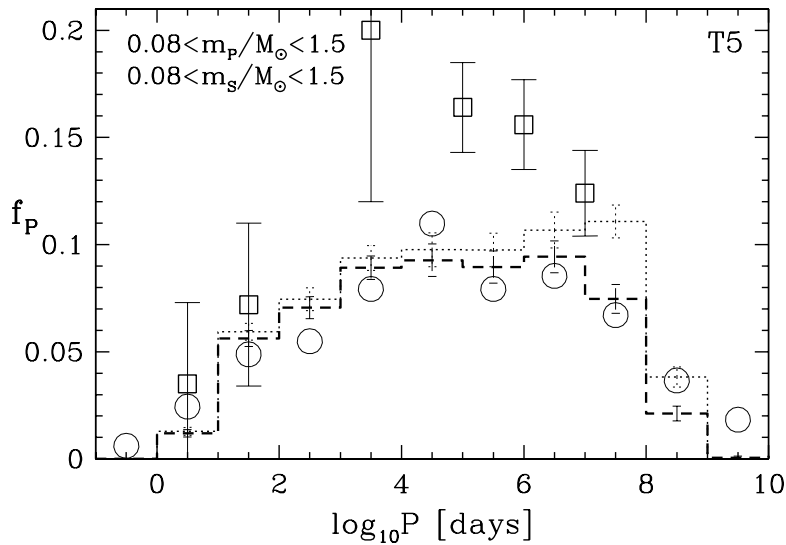}}}
\vskip -60mm
\caption{ The period distribution function of all star--star binaries
in model~T5 which assumes the SMwBDs but has a non-standard IMF in the
BD mass range ($\alpha_0=-0.5$).  The thin dotted histogram is the
initial distribution, while the dashed histogram is the final
distribution at an age of about 15~Myr.  The histograms are averages
of 140~renditions of model~T5, and the error bars are standard
deviation of the mean values.  The open squares are pre-main sequence
data and the open circles are main-sequence data as in
Fig.~\ref{fig:fp1}. }
\label{fig:fpT5}
\end{center}
\end{figure}
\begin{figure}
\begin{center}
\rotatebox{0}{\resizebox{0.6 \textwidth}{!}
{\includegraphics{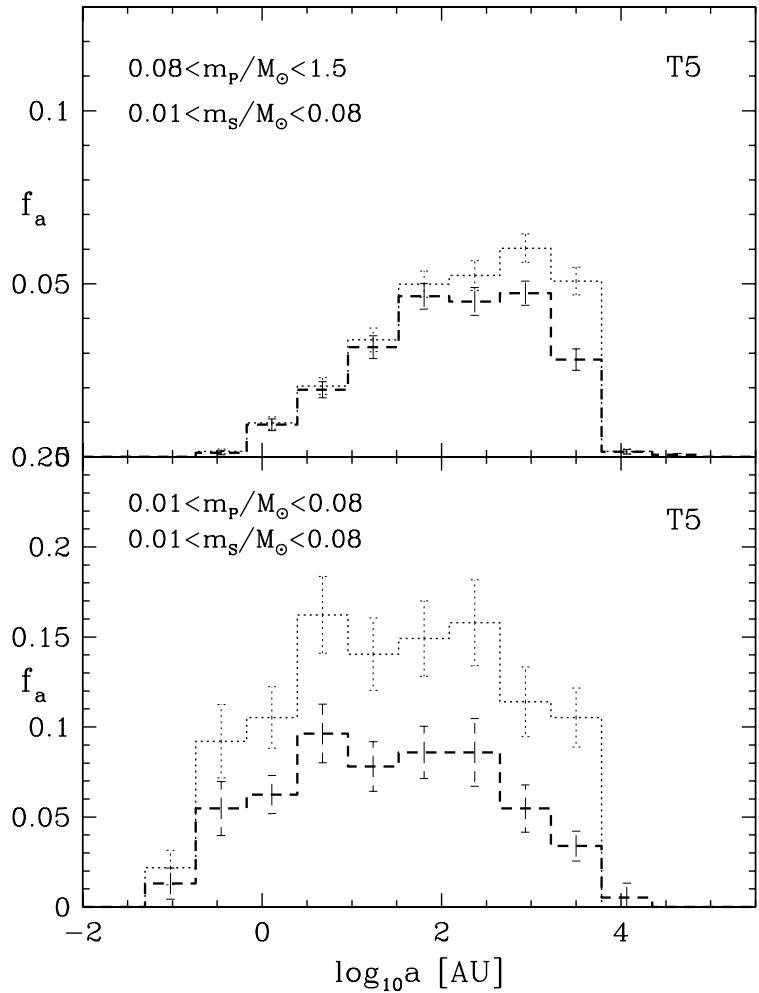}}}
\vskip 0mm
\caption
{{\bf Upper panel:} The distribution of semi-major axes of BD
companions to all stellar primaries according to model~T5 which has a
non-standard IMF in the BD mass range.  {\bf Lower panel:} The
distribution of semi-major axes of all BD--BD binaries in model~T5.
In both panels the thin dotted histograms are the initial
distributions, while the dashed histograms depict the final ($\approx
15$~Myr old) distributions.  All histograms are averages of 140~model
renditions, and the error bars are standard deviation of the mean
values.}
\label{fig:semiT5}
\end{center}
\end{figure}

 
However, the distribution of periods of star--star binaries is still
inconsistent with the pre-main sequence data for TA (the open squares
in Fig.~\ref{fig:fpT5}), although the discrepancy is reduced relative
to models~T0 and~T1 as expected.  The number of BD companions to
stellar primaries with separations in the range $10^2-10^4$~AU in the
Briceno et al. (2002) survey is expected to be about
$0.13\times60\approx8$ systems (Fig.~\ref{fig:semiT5}), whereas one
has been detected.  The model cannot be excluded with confidence
higher than the three-sigma level using this argument.  Similarly,
there should be approximately $0.17\times 10\approx2$ BD--BD binaries
with separations in the same range. This is consistent with a null
detection.

In conclusion, model~T5 which assumes random pairing from a
non-standard IMF which is depleted in BDs relative to the standard IMF
such that the observed number of BDs per star is still consistent with
the Briceno et al. (2002) result, leads to improved agreement with the
star--star binary data, but still produces too few stellar binaries
with periods in the range $10^{2.5}\simless P/{\rm d} \simless
10^{6.5}$. The number of star--BD binaries is only marginally
consistent with the data, while the number of BD--BD binaries (two) is
consistent with the data (zero).

The situation for star--star binaries would improve further if $-1.5 <
\alpha_0 < -0.5$, which is still consistent with the data of Briceno
et al. (2002) within two sigma as is evident in Fig.~\ref{fig:r}.
This can be estimated without doing additional $N$-body calculations
by noting from Figs.~\ref{fig:fp2} and~\ref{fig:fpT5} that decreasing
$\alpha_0$ by~0.8 implies an increase in $f_{\rm P}$ for star--star
binaries by a factor of about~1.2 in the period range
$10^3-10^7$~d. Decreasing $\alpha_0$ by a further~0.8 to
$\alpha_0=-1.3$ thus implies a further increase in $f_{\rm P}$ by a
factor of about 1.2 to a level of $f_{\rm P}\approx0.11$ for
$10^3\simless P/{\rm d} \simless 10^7$.  Such a model can be discarded
with very high confidence by applying the Wilcoxon Signed-Rank test
(all of the four pre-main sequence data points lie above the model,
Bhattacharyya \& Johnson 1977).  Thus, the intermediate number of
isolated BDs and the very small number of companion BDs implies that
random pairing out of any IMF is impossible since dynamical evolution
is almost negligible in TA. A BD-poor IMF would have too few isolated
BDs, while a BD-rich IMF would lead to too many BD companions.

Therefore, {\it in TA the BDs appear to follow different pairing rules
than the stars} by being less frequently in binary systems than stars
are.  It is useful to note that in TA the stars alone can be described
very well by the SM. The SM yields, as the initial star--star period
distribution function, the dotted histogram in Fig.~\ref{fig:fp1}, and
the star---star period distribution function will be eroded only
slightly at long periods in TA-like aggregates, thus being consistent
with the observations in TA.  According to this view, {\it the BDs
would be a population not sharing the same formation history as the
stars}.

\section{Concluding remarks}
\label{sec:concs}

Non-hierarchical multiple systems decay too rapidly leading to too
many single stars and too few binary systems when compared to the
about 1~Myr old pre-main sequence stellar population in TA.
Pre-stellar cloud-core fragmentation therefore appears to mostly form
binary and long-lived hierarchical multiple stellar systems. Available
fragmentation models suggest that unfinished embryos can be expelled
from such forming multiple stellar systems.

This may be used to frame hypothesis~A (\S~\ref{sec:hyp}) which leads
to the standard model (SM) of star formation by assuming that
star-formation always produces a stellar population rich in binaries
and a BD population that is added with separate kinematical and binary
properties. This is the basis of our standard model without BDs
(\S~\ref{sec:SM}) which has been shown in the past to lead to
excellent agreement with the stellar data in TA and the Galactic field
as well as young clusters, and thus indicates a remarkable degree of
invariance of the star-formation products.

In this contribution we test in detail the alternative hypothesis~B
which assumes that pre-stellar cloud-core fragmentation leads to
stellar and sub-stellar binary systems with universal properties that
are taken from the successful SM. Hypothesis~B is based in part on the
observational evidence by Briceno et al. (2002) and White \& Basri
(2003) who suggest that BD-formation cannot be distinguished from
stellar-formation on the basis of their accretion properties and their
kinematics and spatial distribution.  The resulting standard model
with BDs (SMwBDs, \S~\ref{sec:SMwBDs}) is tested here against
observational data that include the number of BDs per star detected in
TA and the ONC as well as the form of the IMFs and of the period
distribution functions of binary systems.  With this goal in mind
stellar-dynamical models are evolved to yield theoretical data that
can be compared with the empirical data.

Based on their data, Briceno et al. (2002) conclude that the IMF of
BDs in TA differs from the observed IMF in the central part of the ONC
by having significantly fewer BDs per star.  In contradiction to the
claim by Briceno et al., the SMwBDs with the standard IMF yields
excellent agreement with the observed number of BDs per star in TA and
the ONC (Fig.~\ref{fig:r}). The reason for this is that dynamical
evolution of the TA groups is very mild leading to BDs staying
locked-up in binary systems, while in the ONC a large fraction of the
primordial binary population is disrupted freeing BDs.  {\it The
observed number of BDs per star is therefore not a reliable measure of
IMF differences if used alone}. The properties of binary systems need
to be consulted as well to allow more robust conclusions. This is true
for any population.

Furthermore, the SMwBDs is in good agreement with the observed period
distribution function of star--star binaries in the ONC and the
Pleiades. The observed IMFs are also consistent with the standard IMF
in both clusters and in TA (\S~\ref{sec:stand}). Thus, in both the ONC
and the Pleiades, the data are consistent with the BDs having formed
with the same properties like the stars and therefore with the
scenario that the fragmentation of pre-stellar cloud-cores extends to
sub-stellar core masses and produces uncorrelated components, and that
the orbital elements of the resulting binary do not depend on the mass
of the primary.  The prediction of the number of star--BD and BD--BD
binaries as a function of semi-major axis is given in
Fig.~\ref{fig:fpAB_BDs} for future tests of the SMwBDs in the Pleiades
and the ONC. 

However, available Galactic-field data on the binary proportion near
and below the sub-stellar mass limit already pose a problem for the
SMwBDs, because this model predicts the binary fraction to be much
higher than is observed.  There is also a marked disagreement between
the SMwBDs and the data in TA. The model produces too few star--star
and too many star--BD binaries.  The attempt to save the SMwBDs for TA
by changing the BD IMF fails. Random pairing produces too many
star--BD binaries for all BD IMFs that are consistent with the
observed number of BDs per star by Briceno et al. (2002).

Hypothesis~B thus leads to a contradiction with the observational data
in TA, and with the properties of Galactic-field binaries near and
below the sub-stellar mass limit.  Hypothesis~B should therefore be
discarded {\it with the implication that BDs may form a distinct
population with pairing properties different to those for stars}, as
is the case in the SM (hypothesis~A).

Returning to the standard model without BDs, it is useful to reiterate
here that this SM leads to excellent agreement with the binary-star
data in TA, the ONC, the Pleiades as well as the Galactic field, and
with the IMF in all four populations.  Hypothesis~A cannot be
rejected, and actually accounts very well for a large range of stellar
populations.

If we were to insist that this is the more appropriate description,
then we would have to infer that {\it the BDs do not have the same
formation history as stars and that they therefore form an additional
population}.  The ONC {\it may be} sporting a higher BD production
efficiency per star than TA, according to the data of Briceno et al.
{\it The IMF of BDs therefore may depend on environment} (as suggested
by Briceno et al.), but this IMF would not be a trivial extension of
the stellar IMF to BDs, as is assumed to be the case in the
SMwBDs. However, before the conclusion can be reached with confidence
that the BD IMF is variable it is necessary to study if loss of BDs
from the shallow potential well of the TA aggregates (Bouvier et
al. 2001) may not account for the observed differences. 

The conclusion that BDs may be treated as an additional population
with its own formation history appears to be in conflict with the
hypothesis that the turbulence spectrum of molecular clouds is
responsible for the mass distribution of pre-stellar cores which is
expected to be continuous across the stellar/sub-stellar mass boundary
(Padoan \& Nordlund 2002). The observation of Motte et al. (1998) that
the pre-stellar cloud core mass spectrum already looks like the
standard IMF supports this notion.  Also, the fragmentation of
collapsing cores is probably a process that cannot depend on the mass
of the final outcome (star or BD) to the degree suggested here.  It is
difficult to see how BDs can come as an entirely separate population,
since the hydrogen burning process occurs much later than the
fragmentation process. On the other hand, the physics of cloud-core
collapse and its fragmentation is far from being understood, so that
it can be argued that the present finding that BDs seem to behave
differently than stars may be allowing the type of insights we have
been hoping for all along. It may be that cloud cores only fragment
when they have sufficient mass, $m_c$, and that the gas reservoir is
always much larger than the mass in the initial fragments. This has
been assumed to be the case by DCB who adopt $m_c\ge 0.25\,M_\odot$
because they require their five initial seeds to have masses larger
than the opacity limit for fragmentation.

Whatever the physical mechanisms may be which ultimately produce the
BD population, the conclusion that BDs may form a population which
does not have binary properties that are a natural extension of
stellar binaries has also been found to be the case for very-low-mass
solar-neighbourhood stars (\S~\ref{sec:dat}) and is evident in
Fig.~\ref{fig:fpm}. Close et al. (2003) find it ``very hard to explain
the total lack of systems with separations larger than 16~AU by
scaling the observed semi-major axis distribution of T~Tauri
stars''. Such a scaling would overproduce the number of wide systems
very significantly compared to observations.  The SMwBDs is,
essentially, such a scaling.

An ``extra BD population'' could be the result of any one of the
following processes: BDs may be ejected unfinished embryos from
accreting systems, or they are hydrostatic cores that loose their
accretion envelopes due to encounters with other proto-stars in rich
clusters, or by photo-evaporation of their accretion envelopes through
nearby O~stars.  These processes will be active for stars so that an
(unknown) fraction of very-low-mass and low-mass stars will probably
also belong to such an extra population.  The above-mentioned
scenarios for the origin of BDs possibly imply that a larger fraction
of hydrostatic cores may be able to accrete the available mass
reservoir and thus become stars in quiescent star forming regions such
as TA, thus perhaps naturally leading to the deduced smaller number of
BDs per star there.  We note that the putative extra population would
need to have accretion and kinematical properties that are consistent
with the observations of Briceno et al. (2002) and White \& Basri
(2003): we recall from \S~\ref{sec:intro} that these authors had
rejected the embryo ejection hypothesis, which we have now returned to,
given the results of the present study.  The possibilities for the
origin of BDs are studied in more detail in Kroupa \& Bouvier (2003b).

In summary and to answer the question posed in the title of this
paper: The results presented here and in other research based on the
SM appear to suggest that {\it the outcome of star-formation is rather
surprisingly invariant}. Specifically, TA, the ONC, the Pleiades and
the Galactic field appear to have had the same initial stellar
population which can be described very well by the SM plus an
additional, primarily single, BD population.  Evident differences can
be attributed to stellar-dynamical evolution, and to the limited
molecular cloud masses which naturally lead to a smaller upper mass
limit in TA than in the ONC.  Only in the sub-stellar mass regime may
the observations indicate different MFs in TA and the ONC, as is in
fact suggested by Briceno et al. (2002). 

\section*{acknowledgements} 
We thank Richard Durisen for useful comments that helped to improve
the manuscript.  PK thanks the staff of the Observatoire de Grenoble
for their very kind hospitality and the Universit\'e Joseph Fourier
for supporting a very enjoyable and productive stay during the summer
of 2002. This work was partially supported by DFG grant KR1635/4,
and made use of Aarseth's {\sc Nbody6} code.


\vfill 

\end{document}